\documentclass[twocolumn]{aastex631}

\received{****}
\revised{*****}
\accepted{****}

\submitjournal{ApJS}

\shorttitle{Au IV Spectrum}
\shortauthors{Zainab et al.}

\graphicspath{{./}{figures/}}

\begin{document}

\title{Critically Evaluated Atomic Data for Au IV Spectrum \footnote{Released on Oct 01, 2021}}

\author[0000-0002-2801-4717]{Aashna Zainab}
\affiliation{Department of Physics, Aligarh Muslim University, Aligarh, UP 202002, India}

\author[0000-0002-1341-6297]{K. Haris}
\affiliation{Department of Physics, Aligarh Muslim University, Aligarh, UP 202002, India}

\author[0000-0002-9115-590X]{S\'ebastien Gamrath}
\affiliation{Physique Atomique et Astrophysique, Université de Mons, B-7000 Mons, Belgium}

\author[0000-0002-3937-2640]{Pascal Quinet}
\affiliation{Physique Atomique et Astrophysique, Université de Mons, B-7000 Mons, Belgium}
\affiliation{IPNAS, Université de Liège, B-4000 Liège, Belgium}

\author[0000-0003-3856-5042]{A. Tauheed}
\affiliation{Department of Physics, Aligarh Muslim University, Aligarh, UP 202002, India}

\correspondingauthor{K. Haris}
\email{kharisphy@gmail.com, kharis.ph@amu.ac.in}

\begin{abstract}

The spectral investigation of the triply ionized gold (Au IV) has been carried out in the wavelength region of 500--2106 \AA. The gold spectra were photographed at the National Institute of Standards and Technology (NIST, USA) on a 10.7-m normal incidence vacuum spectrograph (NIVS) using a sliding spark source as well as on a 3-m NIVS at the Antigonish laboratory in Canada with a triggered spark source. Our analysis is theoretically supported by the pseudo-relativistic Hartree-Fock (HFR) formalism with superposition of configuration interactions implemented in Cowan’s suite of codes. Radiative transition parameters are also calculated using HFR+CPOL (core-polarization effects) model and multiconfiguration Dirac-Hartree-Fock (MCDHF) approach, and their comparisons are used to evaluate the transition rate data. All the previously reported levels of the $5d^{8}$, $5d^{7}6s$, and $5d^{7}6p$ configurations are confirmed, except one and three are newly established. The missing $^{1}S_{0}$ level of $5d^{8}$ is now established at 55277.8 cm$^{-1}$. A total of 981 observed lines (E1-type) classified to 1031 transitions, including 133 newly identified, enabled us to optimize 139 energy levels. Several astrophysically important transitions, forbidden (M1- and E2-types) lines of $5d^{8}$ and $5d^{7}6s$, are provided with their Ritz wavelengths and radiative parameters. A critically evaluated set of energy levels, observed and Ritz wavelengths along with their uncertainties, transition rates, and uniformly-scaled intensities of Au IV lines have been presented. Also, large scale atomic data to compute opacity of Au IV in the kilonova ejecta have been supplemented in this work.
\end{abstract}




\section{Introduction} \label{sec:introduction}

The high resolution spectra of chemically peculiar stars obtained with the Goddard High Resolution Spectrograph on-board Hubble Space Telescope rekindled the interest in the spectroscopy of high-\emph{Z} elements. The rich spectrum of the Hg-Mn $\chi$ Lupi stars is made up of the spectra of singly- and doubly-ionized elements with empty 4\emph{d}- and 5\emph{d}-sub-shells whose atomic structures are quite complex \citep{Leckrone1993}. To obtain an accurate data-set for these elements, it was proposed that their spectral research must be supported by the energy parameters regularities in their iso-electronic sequence~\citep{wyart1993spectra}. Consequently, \citet{Wahlgren-1995-Chi-Lupi} summarized the elemental abundances of Pt, Au, and Hg in these chemically peculiar stars, and Au abundances were estimated from the observation of a few selected lines of Au II and Au III. The Au I and Au II have simple structures, however, the spectrum of Au III and those of its higher charge stages are very dense, and they overlap each other. In such cases, spectroscopic analysis of adjacent charge states is highly useful and aids the accurate line identification for the spectrum under investigation. Besides, the laboratory studies of heavier elements ($Z\ge54$) including gold have also gained high astrophysical interest after the detection of the gravitational waves by the LIGO–Virgo collaboration as its observations ‘GW170817’ and the followed kilonova ‘AT2017gfo’ support the hypothesis that neutron star mergers (NSM) can account for the nucleosynthesis of these heavy elements by r-process \citep{2017ApJ...848L..12A,Kasen-2017-NSM-Opc,2019PrPNP.107..109K}. To explain and model the observed AT2017gfo spectra, large sets of calibrated atomic data are needed for elements beyond iron-peak ($Z>30$) including data on their ionization species (II--V). In this regard, large-scale atomic data for computing the bound-bound opacities of r-process ejecta elements have been given by \citet{Kasen-2017-NSM-Opc,Tanaka-2018-KNe-Prop,Fontes-2020-KNe-Opac,Tanaka-2020-KNe-opac}, however, data for neutral gold (Au) and its ions (Au II--IV) were only included in the work of \citet{Tanaka-2020-KNe-opac}. Recently, \citet{Bromley2020Au1-2} and \citet{McCann2022-Au1-3} have provided more calibrated atomic data for Au I–III ions, and \citet{Gillanders2021-Au-Pt-AT2017gfo} used these data together with data on Pt I–III to estimate Au/Pt abundances in the ejecta of AT2017gfo for photospheric and nebular phases. For photospheric modeling, data on allowed electric-dipole transitions could be important, whereas, \citet{Gillanders2021-Au-Pt-AT2017gfo} had manifested that the accurate data on radiative parameters of forbidden lines are important to interpret the observed kilonova AT2017gfo spectra of late-epochs, $t\ge$~+3~d and/or of NSM ejecta with low temperatures and densities. Besides these, transition rates of electric and magnetic dipole transitions in Au IV were reported by \citet{Taghadomi-2022-Au4-Os1} using GRASP code calculations. These authors also showed the expansion opacity of Au IV for kilonova ejecta at time t~=~1~day and for temperature T = 3700 K and density $\rho$ = 10$^{-13}$ g/cm$^{3}$.

In terms of astrophysical observations, the presence of gold (Au I-II) has been confirmed in the environments of various astrophysical objects, however, so far only its allowed transitions are detected and no forbidden lines are observed yet in the spectra of astrophysical objects \citep[see details in][]{Wahlgren-1995-Chi-Lupi,Gillanders2021-Au-Pt-AT2017gfo}. It should be noted that at present radiative parameters of forbidden transitions are reported only for Au I--III spectra, and no such data are available for the Au IV. Also, our survey on forbidden spectrum of Au IV shows that its several lines fall in the visible-IR (0.5 to 28 $\mu$m) operating region of the recently launched James Webb Space Telescope (JWST), and the role of the JWST will be instrumental for the future kilonovae observations in the IR region \citep{Bartos-2016-JWST-GW} as similar to observations made by the Spitzer Space Telescope for the ‘GW170817’ NSM event \citep{Villar-2018-Spitzer-GW,Wu-2019-Heavy-Elements-GW,Kasliwal-2022-Spitzer-GW}. Therefore, from laboratory to astrophysics, accurate data on energy levels, wavelengths, intensities, and transition rates are greatly needed for various applications ~\citep{Wahlgren-2011-IR-data-JWST}, including estimating the abundances of r-process elements in the kilonova ejecta~\citep{Vieira-2023-Abund-KNe}. These data are also used for improving the accuracy of theoretical calculations of complex atoms/ions. Thus, the present work aims to provide critically evaluated spectral data for triply-ionized gold (Au IV) belonging to the 5\emph{d}-series elements.

Au IV is a member of Os~I iso-electronic sequence. Its ground configuration is [Xe] $4f^{14}5p^{6}5d^{8}$, which consists of 9 energy levels; $^3F_{4,3,2}$, $^3P_{2,1,0}$, $^1D_{2}$, $^1G_{4}$ and $^1S_{0}$ among which $^3F_{4}$ is the lowest. The regular excited configurations, formed by the excitation of one electron from the $5d$ sub-shell, are of the type $5p^{6}5d^{7}$\emph{nl} $(\emph{n}\geq5, \emph{l}\geq0)$. Promotion of two or three electrons leads to the complex configurations such as $5d^{6}6s^{2}$, $5d^{6}6p^{2}$, $5d^{6}6s6p$, and $5d^{5}6s^{2}6p$. These configurations are incorporated in the theoretical calculations for better accuracy.

The first spectral analysis which provided an insight into the energy level structure of Au IV was performed by~\citet{joshi1991Au4}. They studied the $5d^{8}$--$5d^{7}6p$ transition array in the wavelength region of 554--850~\AA.~\citet{wyart1994Au4} provided an extended analysis of Au IV spectrum by identifying about 700 lines of the $(5d^{8}+5d^{7}6s)$--$5d^{7}6p$ transition array in the wavelength region of 554--2080~\AA. They established 33 (out of 38) energy levels of the $5d^{7}6s$ configuration and increased the known levels of $5d^76p$ from 64 to 95. Apart from these, the $4f$ and $5p$ photoabsorption spectra of Au IV have been studied in the wavelength region of 98 to 180 \AA~using a dual laser-plasma technique \citep{Su-2009-photoabs-Au3-5}. This work is beyond the scope of our investigation, therefore, we did not include it in this critical evaluation.

In this work, we have critically evaluated all the previously reported data on the Au IV using the spectrum recorded on a 10.7-m normal incidence vacuum spectrograph, and also by providing a theoretical support for our observations within the pseudo-relativistic Hartree-Fock approach, and the effects of core-polarization are accounted for computing the accurate transition rates for Au IV spectrum. We aim to provide the wavelengths of observed lines (together with their Ritz counterpart), the optimized energy levels with their uncertainties, uniformly-scaled intensities, and transition rates for observed and possibly observable lines of both allowed and forbidden types. The latter types are intra- and inter-configurations transitions between the levels of $5d^{8}$ and $5d^{7}6s$, however, may have implications in astrophysics, in particular to interpret the visible and near-infrared parts of the kilonova spectra associated with NSM event \citep{Gillanders2021-Au-Pt-AT2017gfo}. 

\section{Experimental Details} \label{sec:experiment}

The spectral plates used for this work were the same as those used for the previous analysis of Au III by \citet{Zainab-2019-Au3}. Nevertheless, the necessary details regarding the spectrograms and its recording are provided here for the completeness. The recording of the gold spectra, in the region of 500--2106 \AA, was performed on a 10.7-m normal incidence vacuum spectrograph (NIVS) at the National Institute of Standard and Technology (NIST, USA) using sliding spark as the excitation source. The spectrograph was equipped with a 1200 lines per mm grating which had the first order reciprocal dispersion of 0.78~\AA~mm$^{-1}$. The gold spectra were also recorded at the Antigonish laboratory on a 3-m NIVS with reciprocal dispersion of 1.385~\AA~mm$^{-1}$ using a triggered spark source. The latter was used just for the purpose of ionization separation. Kodak Short Wave Radiation (SWR) plates were used for recording of the spectra. Several tracks of exposures were taken on each plate by varying the experimental conditions, which were achieved by either adjusting the capacitor’s charging potential or by inserting inductor coils in series with the discharge circuit. The variation in line intensity at different discharge settings allowed for a satisfactory ionization separation. The track with the strongest Au IV lines was used to measure the wavelengths and the remaining tracks on each plate were used to accomplish the ionization separation. The relative positions of the spectral lines along with their relative visual-intensities were measured on a Zeiss Abbe comparator at Aligarh Muslim University, Aligarh. The calibration of the spectrograms were then carried out using reference lines of C II--IV, Si II--IV, Al II--III, O II--V~\citep{nist} together with some Au II lines~\citep{rosberg1997Au2}. The general estimate of wavelength uncertainty is due to the combined effect of the statistical deviation of the line position measured on the comparator and the systematic uncertainty of reference wavelengths used in the fitting. The final wavelength uncertainty for sharp and unperturbed lines is estimated to be better than 0.005~\AA~and the figure has been doubled for all perturbed lines.

The spectral lines' intensities were visual estimates representing the photographic emulsion blackening. To bring these intensities on a uniform scale, intensity modelling was performed (see Section. \ref{sec:intensity}).

\section{Analysis of the spectrum} \label{sec:analysis}

The first spectroscopic analysis of Au IV was carried out by~\citet{joshi1991Au4} using the spectra recorded on a 6.65-m NIVS at the Zeeman Laboratory and on a 3-m NIVS at the Antigonish laboratory in the wavelength region of 554--850~\AA. For both the cases, triggered spark was used as the excitation source. Several exposures were recorded on Kodak SWR plates with different experimental conditions. The relative positions of spectral lines were measured on Cospinsca or Grant semiautomatic comparators and the line intensity is reported on a scale of 1 to 99. The wavelength calibration was done using the internal standards of carbon, oxygen, and silicon ions' lines and their measured wavelength uncertainty was 0.005~\AA. On the basis of the classification of 196 lines, 8 out of 9 levels of $5d^{8}$ and 64 out of 110 levels of $5d^{7}6p$ configuration were reported by them.

In the extended analysis of Au IV spectrum,~\citet{wyart1994Au4} reported the lines belonging to $5d^{7}6s$--$5d^{7}6p$ transition array in the region of 1020--1930~\AA. The spectrograms were recorded on a 10.7-m NIVS of the National Bureau of Standards (now known as NIST, USA) having 0.78~\AA~mm$^{-1}$ as the reciprocal dispersion. A sliding spark was used as the excitation source which was operated at different experimental conditions to achieve ionization separation. The spectrograms were measured on a semi-automatic comparator at Observatoire de Paris-Meudon and were calibrated using the reference standards of Cu (copper hollow cathode spectrum was recorded on the same plates), Si, C, N and Al. The accuracy of this measurement was quoted $\pm$0.007 \AA. Additional gold plates, recorded by~\citet{joshi1991Au4} on the same experimental setup in the wavelength region of 554--2080 \AA, were also supplemented to support the analyses. In the $5d^{7}6s$--$5d^{7}6p$ transition array, a total of 657 lines and between $5d^{8}$--$5d^{7}6p$ configurations 47 newly classified lines have been reported by them. This lead to the establishment of 33 out of 38 energy levels of $5d^{7}6s$ configuration and among $5d^{7}6p$ levels, two were revised and 31 being newly established.

Our investigation started with the verification of the previously reported data on Au IV. In total 139 energy levels belonging to the $5d^{8}$, $5d^{7}6s$, and $5d^{7}6p$ configurations have been confirmed. A total of 981 observed lines classified to 1031 transitions were used to obtain the optimized energy levels of Au IV spectrum. All classified lines of Au IV are assembled in Table \ref{tab:lines} with their observed wavelengths and uncertainties. The optimized energy levels of Au IV are given in Table \ref{tab:energy}. The major findings of our present analysis are summarized below:

One level of $5d^{7}6p$ configuration with $J$ = 2, reported earlier by \citet{joshi1991Au4} at $\approx$ 188525 cm$^{-1}$ is now revised to 188170.6 cm$^{-1}$ with the aid of five transitions. We are also successful in establishing the missing $^{1}S_{0}$ level of the $5d^{8}$ configuration. This level was already known in the Hg V \citep{wyart1993Hg5}, Tl VI \citep{raassen1994Tl6}, Pb VII \citep{raassen1994Pb7}, and Bi VIII~\citep{kildiyarova1995Bi8} ions, thus it was possible to recalculate {all the Slater} parameters of this configuration and enhance the accuracy of its predicted value along the iso-electronic sequence. Therefore, after eliminating the previously classified lines in the expected region, the level was established at 55277.8 cm$^{-1}$ and is supported by six Au IV transitions. This energy level value compares well with 55391 cm$^{-1}$, which was predicted by~\citet{wyart1993spectra} based on the orthogonal operator treatment. The determination of the $5d^{8}$ $^{1}S_{0}$ level resulted in new calculation parameters for the ground configuration (see Section \ref{sec:HFR}). Additionally, two new levels of $5d^{7}6p$ configuration with $J$~=~0 were established at 140590.70 cm$^{-1}$ and 149963.62 cm$^{-1}$ on the basis of four and three transitions, respectively. These observed lines were weak in our spectrograms, but they exhibited a distinct Au IV character {similar to that of other previously classified Au IV lines that appeared in these spectrograms}.

\begin{rotatetable*}
\centerwidetable
\begin{deluxetable*}{lllclllclccccc}
\tablecaption{Classified Lines of Au IV~\label{tab:lines}}
\tabletypesize{\scriptsize}
\tablehead{
\colhead{\emph{I}$_{obs}$~$^{a}$} & \colhead{Char.$^{a}$} & 
\colhead{$\lambda_{obs}$ $^{b}$} & \colhead{Unc.$^{b,c}$} & 
\colhead{Lower~Level$^{d}$} & \colhead{Upper~Level$^{d}$} &
\colhead{$\lambda_{Ritz}$ $^{b}$} & \colhead{Unc.$^{b}$} &
\colhead{$\lambda_{prev}^{corr}$ $^{e}$} &
\multicolumn{2}{c}{\emph{A}-value (in s$^{-1}$) $^{f}$} &
\colhead{Acc.$^{g}$} &
\colhead{Line Ref.$^{h}$} &
\colhead{Comm.$^{i}$} \\
\cline{10-11}
\colhead{(arb. u.)} & \colhead{} &
\colhead{(\AA)} & \colhead{(\AA)} &
\colhead{} & \colhead{} &
\colhead{(\AA)} & \colhead{(\AA)} & \colhead{(\AA)} &
\colhead{HFR+CPOL} & \colhead{HFR} &
\colhead{} & \colhead{} & \colhead{} 
} 
\startdata
& & & & 5\emph{d}$^{8}$ $^{3}$\emph{F}$_{4}$ & 5\emph{d}$^{7}$($^{2}$\emph{Da})6\emph{p} $^{1}$\emph{F}$^{\circ}_{3}$ & 524.6742 & 0.0008 & & 2.51e+06 & 3.52e+06 & E & & \\
& & & & 5\emph{d}$^{8}$ $^{3}$\emph{F}$_{4}$ & 5\emph{d}$^{7}$($^{2}$\emph{F})6\emph{p} $^{1}$\emph{F}$^{\circ}_{3}$ & 530.9707 & 0.0011 & & 5.64e+05 & 5.31e+05 & E & & \\
& & & & 5\emph{d}$^{8}$ $^{3}$\emph{F}$_{4}$ & 5\emph{d}$^{7}$($^{2}$\emph{Db})6\emph{p} $^{3}$\emph{F}$^{\circ}_{3}$ & 535.9620 & 0.0009 & & 4.87e+05 & 4.32e+05 & E & & \\
... & ... & ... & ... & ... & ... & ... & ... & ... & ... & ... & ... & ... & ... \\
4900 & & 608.188 & 0.005 & 5\emph{d}$^{8}$ $^{3}$\emph{P}$_{2}$ & 5\emph{d}$^{7}$($^{2}$\emph{G})6\emph{p} $^{1}$\emph{F}$^{\circ}_{3}$ & 608.1802 & 0.0008 & 608.183 & 4.08e+07 & 5.88e+07 & D & TW,J91 & U \\
1800 & * & 608.883 & 0.016 & 5\emph{d}$^{8}$ $^{3}$\emph{F}$_{4}$ & 5\emph{d}$^{7}$($^{2}$\emph{G})6\emph{p} $^{3}$\emph{F}$^{\circ}_{3}$ & 608.8920 & 0.0008 & 608.887 & 4.18e+07 & 3.49e+07 & D & TW,J91 & \\
1800 & * & 608.883 & 0.013 & 5\emph{d}$^{8}$ $^{3}$\emph{P}$_{1}$ & 5\emph{d}$^{7}$($^{2}$\emph{Da})6\emph{p} $^{3}$\emph{F}$^{\circ}_{2}$ & 608.8844 & 0.0011 & 608.887 & 1.11e+08 & 1.40e+08 & D+ & TW,J91 & \\
940 & & 610.140 & 0.005 & 5\emph{d}$^{8}$ $^{3}$\emph{F}$_{3}$ & 5\emph{d}$^{7}$($^{2}$\emph{P})6\emph{p} $^{3}$\emph{D}$^{\circ}_{2}$ & 610.1429 & 0.0008 & 610.143 & 8.61e+07 & 4.78e+07 & D & TW,J91 & \\
... & ... & ... & ... & ... & ... & ... & ... & ... & ... & ... & ... & ... & ... \\
650 & & 1110.180 & 0.005 & 5\emph{d}$^{7}$($^{2}$\emph{F})6\emph{s} $^{3}$\emph{F}$_{2}$ & 5\emph{d}$^{7}$($^{2}$\emph{H})6\emph{p} $^{3}$\emph{G}$^{\circ}_{3}$ & 1110.1858 & 0.0022 & & 6.38e+07 & 8.22e+07 & D+ & TW & \\
1500 & & 1110.657 & 0.005 & 5\emph{d}$^{7}$($^{2}$\emph{G})6\emph{s} $^{3}$\emph{G}$_{4}$ & 5\emph{d}$^{7}$($^{2}$\emph{G})6\emph{p} $^{1}$\emph{H}$^{\circ}_{5}$ & 1110.6593 & 0.0016 & 1110.659 & 8.62e+07 & 8.77e+07 & D+ & TW,W94 & V \\
930 & d & 1110.872 & 0.010 & 5\emph{d}$^{7}$($^{4}$\emph{P})6\emph{s} $^{5}$\emph{P}$_{2}$ & 5\emph{d}$^{7}$($^{2}$\emph{Db})6\emph{p} $^{3}$\emph{D}$^{\circ}_{3}$ & 1110.8667 & 0.0013 & 1110.866 & 2.51e+08 & 2.76e+08 & D+ & TW,W94 & \\
930 & d & 1110.951 & 0.010 & 5\emph{d}$^{7}$($^{2}$\emph{H})6\emph{s} $^{3}$\emph{H}$_{6}$ & 5\emph{d}$^{7}$($^{2}$\emph{H})6\emph{p} $^{3}$\emph{H}$^{\circ}_{5}$ & 1110.9557 & 0.0024 & 1110.951 & 1.55e+08 & 1.42e+08 & D+ & TW,W94 & \\
... & ... & ... & ... & ... & ... & ... & ... & ... & ... & ... & ... & ... & ... \\
480 & SA & 1510.878 & 0.005 & 5\emph{d}$^{7}$($^{4}$\emph{F})6\emph{s} $^{3}$\emph{F}$_{4}$ & 5\emph{d}$^{7}$($^{4}$\emph{F})6\emph{p} $^{5}$\emph{G}$^{\circ}_{4}$ & 1510.8796 & 0.0022 & 1510.880 & 1.31e+08 & 1.25e+08 & D+ & TW,W94 & \\
480 & d & 1511.011 & 0.010 & 5\emph{d}$^{7}$($^{4}$\emph{F})6\emph{s} $^{5}$\emph{F}$_{1}$ & 5\emph{d}$^{7}$($^{4}$\emph{F})6\emph{p} $^{5}$\emph{F}$^{\circ}_{2}$ & 1511.010 & 0.003 & 1511.005 & 8.54e+07 & 1.11e+08 & D+ & TW,W94 & \\
450 & & 1511.442 & 0.005 & 5\emph{d}$^{7}$($^{2}$\emph{P})6\emph{s} $^{3}$\emph{P}$_{1}$ & 5\emph{d}$^{7}$($^{2}$\emph{P})6\emph{p} $^{3}$\emph{D}$^{\circ}_{1}$ & 1511.4414 & 0.0023 & 1511.443 & 9.90e+07 & 1.09e+08 & D+ & TW,W94 & U \\
... & ... & ... & ... & ... & ... & ... & ... & ... & ... & ... & ... & ... & ... \\
89 & & 1891.924 & 0.005 & 5\emph{d}$^{7}$($^{2}$\emph{P})6\emph{s} $^{1}$\emph{P}$_{1}$ & 5\emph{d}$^{7}$($^{4}$\emph{F})6\emph{p} $^{5}$\emph{F}$^{\circ}_{2}$ & 1891.930 & 0.003 & 1891.928 & 3.80e+06 & 5.58e+06 & D & TW,W94 & T \\
... & ... & ... & ... & ... & ... & ... & ... & ... & ... & ... & ... & ... & ... \\
& & & & 5\emph{d}$^{7}$($^{2}$\emph{F})6\emph{s} $^{3}$\emph{F}$_{3}$ & 5\emph{d}$^{7}$($^{4}$\emph{F})6\emph{p} $^{5}$\emph{D}$^{\circ}_{4}$ & 3624.831 & 0.013 & & 2.42e+04 & 2.27e+04 & E & & \\
\enddata
\tablecomments{(A few columns are omitted in this condensed sample, but their footnotes (c, d, f) are fully retained for guidance regarding their types and content.)\\
$^{a}$ Averaged relative observed intensities in arbitrary units are given on a uniform {scale} corresponding to the Boltzmann populations in a plasma with an effective excitation temperature of 5.5 eV, for our spectrum (see Section \ref{sec:intensity}). The intensity value is followed by the line character encoded as follows: bl--blended by other lines; d--diffused; h--hazy; m--masked by other nearby lines; q--asymmetric; s--shaded; SA--self-absorbed; SL--shoulder line of long wavelength side; SS--shoulder line of short wavelength side; w--wide; ++--too intense than expected; ?--questionable line; *--intensity shared by more than one lines.\\
$^{b}$ Observed and Ritz wavelengths are in vacuum for $\lambda$ $<$ 2000 \AA~and in standard air outside this limit. Conversion between air and vacuum was made with the five-parameter formula from \citet{Peck1972index}. Assigned uncertainty of given observed wavelength or computed uncertainty of Ritz wavelength are determined in the level optimization procedure.\\ 
$^{c}$ Observed wavenumber (in vacuum) and its uncertainty. These two columns are omitted in this condensed table.\\ 
$^{d}$ Level designations and their energies from Table \ref{tab:energy}. Two columns for energies are dropped from this condensed table.\\
$^{e}$ Previous observed wavelength, reported either by \citet{joshi1991Au4} or by \citet{wyart1994Au4}, was corrected using Ritz wavelength of this work (see Section \ref{sec:optimization}).\\
$^{f}$ Transition rates or \emph{A}-values, from HFR+CPOL and HFR calculations. The quantity \emph{$|$CF$|$}$_{min}$ is the minimum of their absolute cancellation factors (see Section \ref{sec:Tp-eval}). This column of \emph{$|$CF$|$}$_{min}$ is omitted in this condensed table.\\
$^{g}$ Accuracy code of the \emph{A}-value is given in Table \ref{tab:tp-code}
(see Section \ref{sec:tp-e1}).\\
$^{h}$ Line references: J91-- \citet{joshi1991Au4}; W94-- \citet{wyart1994Au4}; TW--This work.\\
$^{i}$ Comments: {T--the variation of individual line intensities used for intensity averaging are within a factor of one; U--those vary within a factor of two; V--those vary within a factor of three; W--those vary by more than a factor of three but less than ten; X--the variation is more than a factor of ten, hence it is excluded from the intensity averaging; Y--the line reported by previous observer is a perturbed line, therefore, excluded from the intensity averaging.}\\
(This table is available in its entirety in machine-readable form.)}
\end{deluxetable*}
\end{rotatetable*}


\begin{rotatetable*}
\centerwidetable
\begin{deluxetable*}{lccclccclclclc}
\tablecaption{Optimized Energy Levels of Au IV\label{tab:energy}}
\tablewidth{800pt}
\tabletypesize{\scriptsize}
\movetabledown=1in
\tablehead{
\colhead{Configuration$^{a}$} & \colhead{Term$^{a}$} & 
\colhead{\emph{J}} & \colhead{Energy} & \colhead{Unc.$^{b}$} & \colhead{Perc.$^{c}$} & \colhead{Perc2$^{c}$} & \colhead{Conf2$^{c}$} &
\colhead{Perc3$^{c}$} & \colhead{Conf3$^{c}$} &
\colhead{NoL$^{d}$} & \colhead{$\Delta$E$_{o-c}^{e}$} 
& \colhead{Comm.$^{f}$} \\
\colhead{} & \colhead{} & 
\colhead{} & \colhead{(cm$^{-1}$)} & \colhead{(cm$^{-1}$)} &
\colhead{} & \colhead{} & \colhead{} &
\colhead{} & \colhead{} & \colhead{} &
\colhead{(cm$^{-1}$)} & \colhead{} 
} 
\startdata
5\emph{d}$^{8}$ & $^{3}$\emph{F} & 4 & 0.00 & 0.20 & 94 & 5 & 5\emph{d}$^{8}$ $^{1}$\emph{G} & & ... & 39 & 131 & \\
5\emph{d}$^{8}$ & $^{3}$\emph{P} & 2 & 6630.12 & 0.18 & 46 & 35 & 5\emph{d}$^{8}$ $^{1}$\emph{D} & 18 & 5\emph{d}$^{8}$ $^{3}$\emph{F} & 43 & -63 & \\
5\emph{d}$^{8}$ & $^{3}$\emph{F} & 3 & 12293.66 & 0.18 & 99 & & ... & & ... & 40 & 41 & \\
5\emph{d}$^{8}$ & $^{3}$\emph{F} & 2 & 18100.97 & 0.16 & 48 & 46 & 5\emph{d}$^{8}$ $^{3}$\emph{P} & & ... & 48 & -37 & \\
5\emph{d}$^{8}$ & $^{3}$\emph{P} & 0 & 18185.2 & 0.4 & 82 & 16 & 5\emph{d}$^{8}$ $^{1}$\emph{S} & & ... & 11 & -161 & \\
5\emph{d}$^{8}$ & $^{3}$\emph{P} & 1 & 21489.95 & 0.21 & 97 & 2 & 5\emph{d}$^{7}$($^{2}$\emph{P})6\emph{s} $^{3}$\emph{P} & & ... & 27 & -9 & \\
5\emph{d}$^{8}$ & $^{1}$\emph{G} & 4 & 25702.13 & 0.16 & 94 & 5 & 5\emph{d}$^{8}$ $^{3}$\emph{F} & & ... & 31 & 33 & \\
5\emph{d}$^{8}$ & $^{1}$\emph{D} & 2 & 30735.67 & 0.20 & 49 & 35 & 5\emph{d}$^{8}$ $^{3}$\emph{F} & 14 & 5\emph{d}$^{8}$ $^{3}$\emph{P} & 30 & -5 & \\
5\emph{d}$^{7}$($^{4}$\emph{F})6\emph{s} & $^{5}$\emph{F} & 5 & 51296.97 & 0.15 & 86 & 13 & 5\emph{d}$^{7}$($^{2}$\emph{G})6\emph{s} $^{3}$\emph{G} & & ... & 12 & -105 & \\
5\emph{d}$^{8}$ & $^{1}$\emph{S} & 0 & 55277.8 & 0.3 & 83 & 15 & 5\emph{d}$^{8}$ $^{3}$\emph{P} & & ... & 6 & 64 & N\\
... & ... & ... & ... & ... & ... & ... & ... & ... & ... & ... & ... & ...\\
5\emph{d}$^{7}$($^{4}$\emph{P})6\emph{s} & $^{5}$\emph{P} & 3 & 68890.98 & & 67 & 17 & 5\emph{d}$^{7}$($^{4}$\emph{F})6\emph{s} $^{5}$\emph{F} & 8 & 5\emph{d}$^{7}$($^{4}$\emph{F})6\emph{s} $^{3}$\emph{F} & 30 & 39 & B\\
5\emph{d}$^{7}$($^{4}$\emph{P})6\emph{s} & $^{5}$\emph{P} & 1 & 72968.57 & 0.09 & 47 & 24 & 5\emph{d}$^{7}$($^{4}$\emph{F})6\emph{s} $^{5}$\emph{F} & 19 & 5\emph{d}$^{7}$($^{2}$\emph{P})6\emph{s} $^{3}$\emph{P} & 21 & 133 & \\
5\emph{d}$^{7}$($^{2}$\emph{G})6\emph{s} & $^{3}$\emph{G} & 5 & 75093.20 & 0.10 & 59 & 23 & 5\emph{d}$^{7}$($^{2}$\emph{H})6\emph{s} $^{3}$\emph{H} & 12 & 5\emph{d}$^{7}$($^{4}$\emph{F})6\emph{s} $^{5}$\emph{F} & 18 & -165 & \\
5\emph{d}$^{7}$($^{4}$\emph{F})6\emph{s} & $^{3}$\emph{F} & 3 & 75819.28 & 0.08 & 58 & 14 & 5\emph{d}$^{7}$($^{4}$\emph{P})6\emph{s} $^{5}$\emph{P} & 13 & 5\emph{d}$^{7}$($^{2}$\emph{Db})6\emph{s} $^{3}$\emph{D} & 30 & -30 & \\
... & ... & ... & ... & ... & ... & ... & ... & ... & ... & ... & ... & ...\\
5\emph{d}$^{7}$($^{2}$\emph{Da})6\emph{s} & $^{1}$\emph{D} & 2 & 123610 & 200 & 75 & 7 & 5\emph{d}$^{7}$($^{2}$\emph{Db})6\emph{s} $^{1}$\emph{D} & 5 & 5\emph{d}$^{7}$($^{2}$\emph{F})6\emph{s} $^{3}$\emph{F} & 0 & & L\\
5\emph{d}$^{7}$($^{4}$\emph{F})6\emph{p} & $^{5}$\emph{D}$^{\circ}$ & 4 & 120217.42 & 0.09 & 42 & 20 & 5\emph{d}$^{7}$($^{4}$\emph{F})6\emph{p} $^{5}$\emph{F}$^{\circ}$ & 13 & 5\emph{d}$^{7}$($^{2}$\emph{G})6\emph{p} $^{3}$\emph{F}$^{\circ}$ & 7 & 104 & \\
5\emph{d}$^{7}$($^{4}$\emph{F})6\emph{p} & $^{3}$\emph{G}$^{\circ}$ & 5 & 122495.68 & 0.21 & 30 & 30 & 5\emph{d}$^{7}$($^{4}$\emph{F})6\emph{p} $^{5}$\emph{F}$^{\circ}$ & 21 & 5\emph{d}$^{7}$($^{4}$\emph{F})6\emph{p} $^{5}$\emph{G}$^{\circ}$ & 5 & 359 & \\
5\emph{d}$^{7}$($^{2}$\emph{P})6\emph{p} & $^{3}$\emph{P}$^{\circ}$ & 1 & 130380.25 & 0.10 & 23 & 12 & 5\emph{d}$^{7}$($^{2}$\emph{Db})6\emph{p} $^{3}$\emph{D}$^{\circ}$ & 10 & 5\emph{d}$^{7}$($^{4}$\emph{P})6\emph{p} $^{5}$\emph{D}$^{\circ}$ & 8 & -145 & \\
... & ... & ... & ... & ... & ... & ... & ... & ... & ... & ... & ... & ...\\
5\emph{d}$^{7}$($^{2}$\emph{Da})6\emph{p} & $^{1}$\emph{P}$^{\circ}$ & 1 & 207500 & 400 & 50 & 20 & 5\emph{d}$^{7}$($^{2}$\emph{Da})6\emph{p} $^{3}$\emph{D}$^{\circ}$ & 6 & 5\emph{d}$^{7}$($^{2}$\emph{Db})6\emph{p} $^{3}$\emph{P}$^{\circ}$ & 0 & & L\\
5\emph{d}$^{7}$($^{2}$\emph{Da})6\emph{p} & $^{3}$\emph{D}$^{\circ}$ & 3 & 207800 & 400 & 46 & 16 & 5\emph{d}$^{7}$($^{2}$\emph{Db})6\emph{p} $^{3}$\emph{D}$^{\circ}$ & 15 & 5\emph{d}$^{7}$($^{2}$\emph{Da})6\emph{p} $^{1}$\emph{F}$^{\circ}$ & 0 & & L\\
\enddata
\tablecomments{\\$^{a}$ The level designations are in LS coupling scheme. Designations in Configuration and Term may correspond to a component with second/third/fourth largest percentage in the eigenvector and they are opted to uniquely define the levels (see text).\\
$^{b}$ The quantity given in column Unc. is the uncertainty of separation from the ``base'' level 5\emph{d}$^{7}$($^{4}$\emph{P})6\emph{s} $^{5}$\emph{P}$_{3}$ at 68890.98 cm$^{-1}$ (see Section \ref{sec:optimization}). To estimate the uncertainties of excitation energies from the ground level, the given values in column Unc. should be combined in quadrature with the uncertainty of the ground level, 0.20~cm$^{-1}$. All uncertainties are derived by the LOPT code.\\
$^{c}$ The first leading percentage value refers to the configuration and term given in the first two columns of the table. The 2nd, 3rd, 4th, and 5th percentages refer to the configuration and term subsequent to them. The four columns (for 4th and 5th percentages and their compositions) are omitted in this condensed table.\\
$^{d}$ Number of observed lines determining the level in the level optimization procedure. Zero for unobserved levels.\\
$^{e}$ Differences between energy observed and those calculated in the parametric least-squares fitting (LSF) with Cowan's code. Blank for unobserved levels.\\
$^{f}$ Comments: B--the base level for presentation of uncertainties; L--the level position and its uncertainty are obtained in the parametric LSF calculations with Cowan's codes; N--new level; R--revised level.}
(This table is available in its entirety in machine-readable form.)
\end{deluxetable*}
\end{rotatetable*}

\section{OPTIMIZATION of ENERGY LEVELS} \label{sec:optimization}

\begin{figure}[t]
\epsscale{1.15}
\plotone{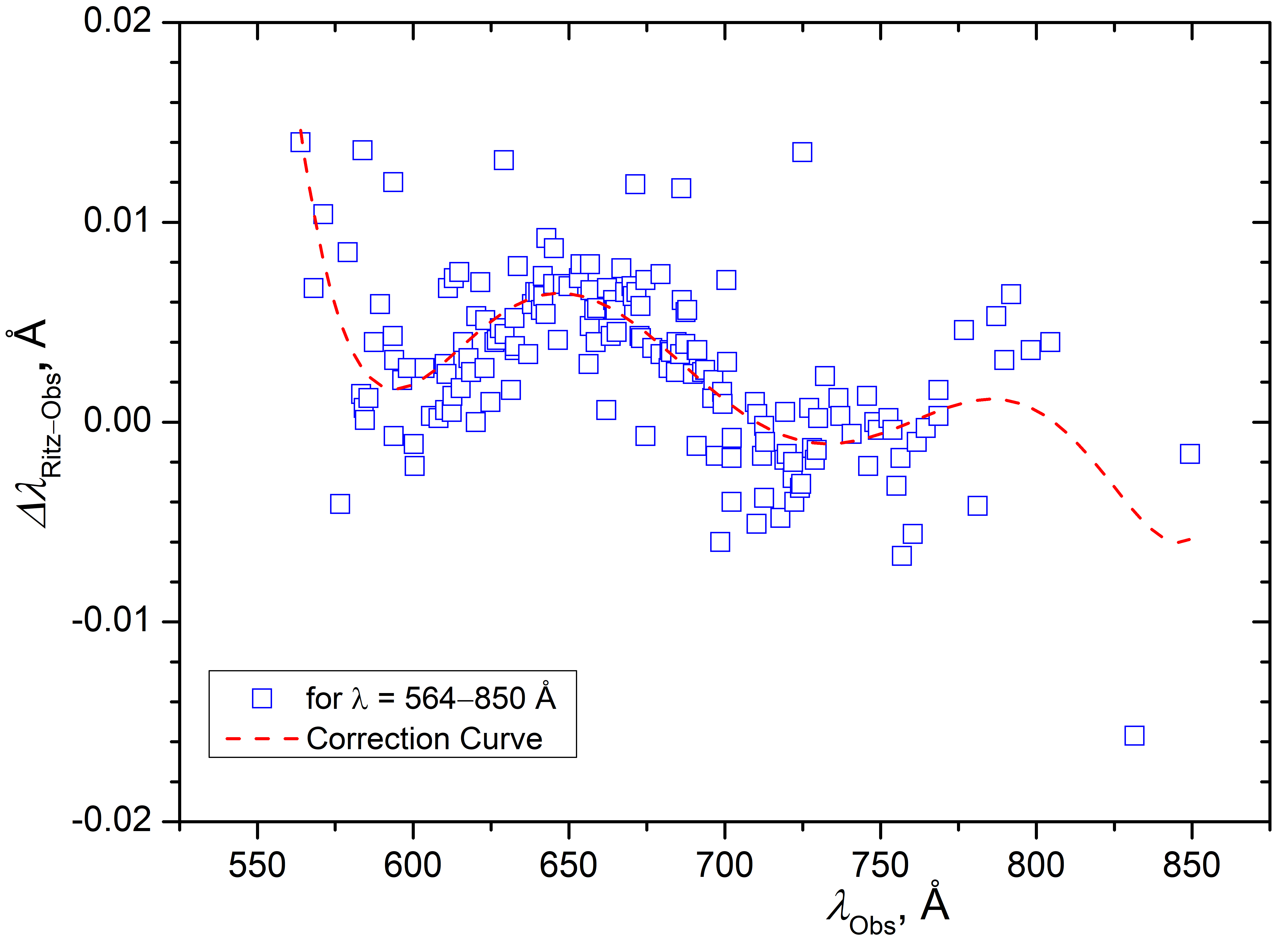}
\caption{Comparison of wavelengths observed by \citet{joshi1991Au4} with those of Ritz in this work. The dashed line represents a correction curve fitted with a seventh order polynomial function (see text).}
\label{fig:J91}
\end{figure}

The least-squares level optimization procedure was done by using the computer code LOPT~\citep{kramida2011lopt}, which helps in obtaining the best fitted energy level values for all observed lines. This process also checks the consistency of the line classifications, by comparing the observed wavelengths with their corresponding Ritz values. A thorough description of this method can be found in several previous publications \citep{kramida2013critical,kramida2013In2,kramida2013Ag2,haris2014Sn2}. The optimization procedure was started by including only the previously classified lines of Au IV which led to an initial estimate of the energy level values and Ritz wavelengths of several possibly observable lines with their uncertainties. These lines were searched out in our line list and the obtained lines (about 130) along with their satisfactory character and intensities were added to the optimization procedure. The entire process involves several iterations of level optimization, consequently, the best optimized energy values and uncertainties were obtained for 139 levels (see Table \ref{tab:energy}) with the help of 981 distinct lines that are classified to 1031 transitions (see Table \ref{tab:lines}). To determine the final level uncertainties given in Table \ref{tab:energy}, the 5\emph{d}$^{7}$($^{4}$\emph{P})6\emph{s} $^{5}$\emph{P}$_{3}$ level at 68890.98 cm$^{-1}$ was selected as a base level, because this level has maximum number of accurately measured transitions.

\begin{figure}
\epsscale{1.16}
\plotone{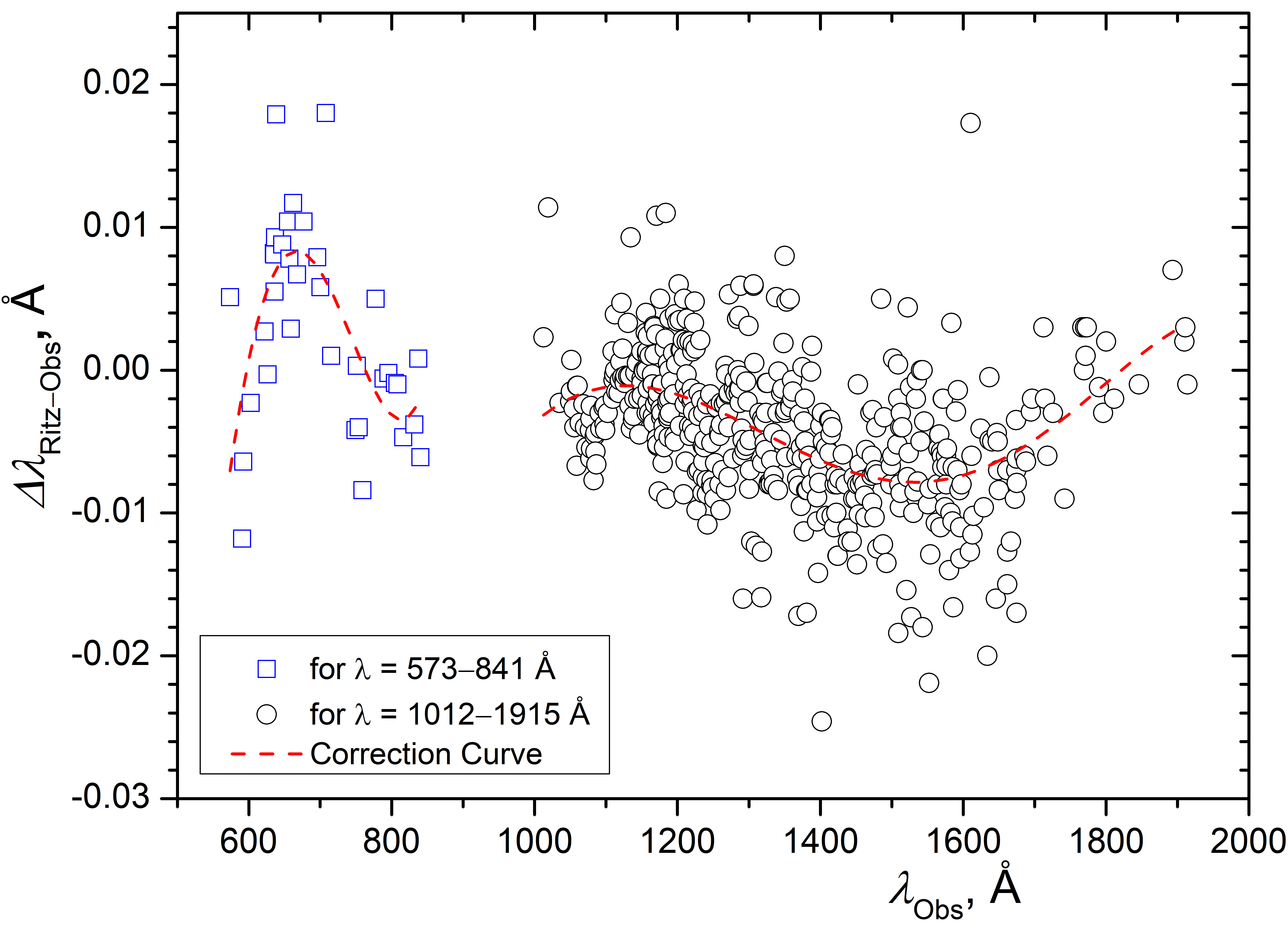}
\caption{Comparison of wavelengths observed by \citet{wyart1994Au4} with those of Ritz in this work. The dashed lines represent the correction curves -- the first ($\lambda$~$<$~1000~\AA) is fitted with a fourth order polynomial and the second is made with a fifth order polynomial function (see text).}
\label{fig:W94}
\end{figure}

Using these results, the final Ritz wavelengths ($\lambda_{Ritz}$) with their estimated uncertainties were computed for all observed and several possibly observable lines (see Table \ref{tab:lines}). These counterpart (i.e., $\lambda_{Ritz}$) wavelengths are at least two-times (or more) accurate than the observed ones, therefore, were served as reference lines to check internal consistencies of previous measurements \citep{joshi1991Au4,wyart1994Au4}. The results of these comparisons are shown in Figure \ref{fig:J91} and \ref{fig:W94}. The suitable systematic correction curves were made with the help of different polynomial fits (see captions of Figure \ref{fig:J91} and \ref{fig:W94}), and the corrected wavelengths of previous measurements were provided in Table \ref{tab:lines}. The agreement between corrected and Ritz wavelengths was found to be satisfactory within a standard deviation of 0.005~\AA, which fairly agrees with their original claimed accuracy.

\section{Reduction of Line Intensities} \label{sec:intensity}

The line intensities play a vital role for the proper identification of the observed lines. Generally, different observers use different scales to express the relative line intensities. One of the widely used techniques to provide the relative intensities is by the visual assessment of blackening of the photographic emulsion. Besides, optical densitometers are also used for this purpose. As the data under study were observed by several authors using different excitation sources and recording equipments, there was a need to convert the overall intensities to a uniform scale. This reduction was achieved by using local thermodynamic equilibrium (LTE) approximation; the detailed procedure of which has been explained in many laboratory spectrum analyses~\citep{kramida2013critical, kramida2013In2,kramida2013Ag2,haris2014Sn2}. In this method, the effective excitation temperature for any particular atom or ion is determined using a simplified LTE model applied to the observation (line intensities). This method requires accurate transition probabilities (g$A$-values) as well, however, they are often provided from computations. Once the approximate effective excitation temperature (\emph{T}$_{exc}$) for the light source used has been determined, the spectral response function(s) of the instrument (including the detector) can easily be determined by comparing the observed and calculated intensities. This function is subtracted from the observed intensities to remove the effects of any spectral variations caused due to the instrument. This leads to the reduction of corrected intensities and then to obtain a more accurate value of~\emph{T}$_{exc}$ from the Boltzmann plot. Several iterations of this procedure is performed until a reasonable convergence is reached. In the same manner \emph{T}$_{exc}$ is derived for each sets of observations. Finally, all of them were effectively reduced to a global uniform scale with a common~\emph{T}$_{exc}$ obtained from one of the observations. If the same set of observed lines are available from multiple sources, the modeled intensities for each of them is averaged out.

The above mentioned procedure was followed for the intensities of Au IV lines observed on our spectrograms, which were taken from seven different plates recorded at different experimental conditions on the same instrument. It should be noted that our originally recorded intensities are visual estimates of photographic blackening of emulsion. The modeling of the intensities was done using the weighted transition rates obtained from HFR+CPOL calculations. The final \emph{T}$_{exc}$ for the light source used for the present study was derived to be 5.5~eV. This temperature was chosen as the base to reduce the intensities of other observations \citep{joshi1991Au4, wyart1994Au4}.

\citet{joshi1991Au4} measured the relative positions of the observed lines on a Cospinsca or Grant semiautomatic comparator and reported the intensity on the scale of 1 to 99. On the other hand, the intensities reported by~\citet{wyart1994Au4} were obtained mostly from a densitometer, which they traced for four separate spectral plates. For a few set of lines the visual estimates were also made. \citet{wyart1994Au4} mentioned that the comparisons of lines' intensities were limited to a small span of wavelength range, and intensities were not consistent for the entire range of observation. He also supplemented the linelist with the gold plates recorded on the same experimental setup as~\citet{joshi1991Au4}. To reduce these \citep{joshi1991Au4, wyart1994Au4} relative intensities into a common \emph{T}$_{exc}$ scale, we combined their observations and a \emph{T}$_{exc}$~=~5.3~eV was obtained for the combined observation. To compare (and to average) these line intensities with that of our spectrograms, the reduced (corrected) intensities were further scaled with respect to the common \emph{T}$_{exc}$~=~ 5.5~eV of our work. In the end, the modelled line intensities for the commonly observed lines were compared, and an averaging of them was made if they agree within a standard deviation of one order of magnitude. Our final modelled line intensities with their line characters (\emph{I}$_{obs}$ \& Char.) are given in the first two columns of Table \ref{tab:lines}, and the lines particularly excluded from the intensity averaging, marked with `X' symbol, are given in the last column of the same table. 
\section{Theoretical Interpretation}\label{sec:theory}

To obtain theoretical support for our experimental findings, the theoretical calculations for energy levels, wavelengths, and transition rates (\emph{A}-values) of Au IV was performed using Cowan’s codes~\citep{Cowan1981-theory-code}. This atomic structure calculation code works on the formalism of the pseudo-relativistic Hartree-Fock (HFR) method with the superposition of interacting configurations. However, it has been shown recently \citep{Quinet2006-Os1-2,Xu2007Ir1-2} that inclusion of the core polarization effect (CPOL) is significantly important to achieve the accurate theoretical radiative lifetimes (or transition~rates) for Os~I isoelectronic sequence. This motivated us to perform a new set of improved HFR+CPOL calculations.

To obtain the radiative lifetime data for the levels of $5d^{8}$ and $5d^{7}6s$ configurations, in addition to HFR and HFR+CPOL calculations, separate computations were also made with a fully-relativistic multiconfiguration Dirac-Hartree-Fock (MCDHF) method \citep{Grant2007-book, Fischer2016-mcdhf} implemented in the latest version of General Relativistic Atomic Structure Program, namely GRASP2018 \citep{Fischer2019-GrASP-2K18}.

Further details of these individual calculations are described in below sub-sections. The results and discussion on the critical evaluation of transition rate data produced by different theoretical methods are given in section \ref{sec:Tp-eval}.

\subsection{The HFR Calculations} \label{sec:HFR}

Our extensive HFR calculations contain all major interacting configurations for both parities. The even parity set consists of $5d^{8}$, $5d^{7}ns$ (\emph{n} = 6--8), $5d^{7}nd$ (\emph{n} = 6--8), $5d^{6}6l^{2}$ (\emph{l} = $s$, $p$), $5d^{6}6s6d$, and $5d^{6}6s7s$ configurations; whereas $5d^{7}np$ (\emph{n} = 6--8), $5d^{7}nf$ (\emph{n} = 5--7), $5d^{6}6snp$ (\emph{n} = 6, 7), and $5d^{5}6s^{2}6p$ configurations were included in the odd parity system. The initial scaling for the average energy $(E_{av})$ and spin-orbit interaction $(\zeta_{nl})$ parameters were kept at 100\% of the HFR values, while the radial wave integral $(F^{k})$, exchange integral $(G^{k})$, and configuration interaction integral $(R^{k})$ were fixed at 75\% of the HFR values. A least-squares parametric fitting (LSF) was performed to minimize the differences between observed and theoretical energy level values. The standard deviations of the LSF calculations for the even and odd parity systems were found to be 92 cm$^{-1}$ and 176 cm$^{-1}$, respectively. The fitted LSF parameters are shown in Table \ref{tab:params}, and were further used to compute the improved transition parameters.

\begin{deluxetable*}{llcccccc}[t]
\tablecaption{LSF parameters for Au IV} 
\label{tab:params}
\tablehead{
\colhead{Configuration$^{a}$} & \colhead{Slater~Parameter$^{a}$} &
\colhead{LSF$^{a}$} & \colhead{Unc.$^{b}$} &
\colhead{Index$^{c}$} & \colhead{HFR$^{a}$} &
\colhead{LSF/HFR$^{a}$} & \colhead{Comments$^{d}$} \\
\colhead{} & \colhead{} & \colhead{(cm$^{-1}$)} &
\colhead{(cm$^{-1}$)} & \colhead{} &
\colhead{(cm$^{-1}$)} & \colhead{} &
\colhead{}
}
\startdata
5\emph{d}$^{8}$ & E$_{av}$ & 18247.6 & 37.0 & & 19163.7 & 0.952 & \\
5\emph{d}$^{8}$ & $F^{2}(5d,5d)$ & 59636.9 & 157.0 & 1 & 73337.2 & 0.813 & \\
5\emph{d}$^{8}$ & $F^{4}(5d,5d)$ & 44803.0 & 275.0 & 2 & 48865.2 & 0.917 & \\
5\emph{d}$^{8}$ & $\alpha(5d)$ & 17.2 & 4.0 & 3 & 0.0 & & \\
5\emph{d}$^{8}$ & $\beta(5d)$ & -437.9 & -74.0 & 7 & 0.0 & & \\
5\emph{d}$^{8}$ & $T(5d)$ & 0.0 & & & 0.0 & & F \\
5\emph{d}$^{8}$ & $\zeta(5d)$ & 5589.7 & 16.0 & 4 & 5769.1 & 0.969 & \\
5\emph{d}$^{7}$6\emph{s} & $E_{av}$ & 85932.6 & 20.0 & & 89638.1 & 0.959 & \\
... & ... & ... & ... & ... & ... & ... & ... \\
... & $R^{k}$ & & & & & 0.750 & R \\
5\emph{d}$^{7}$6\emph{p} & $E_{av}$ & 164844.4 & 21.0 & & 166636.7 & 0.989 & \\
5\emph{d}$^{7}$6\emph{p} & $F^{2}(5d,5d)$ & 60028.6 & 247.0 & 1 & 75500.9 & 0.795 & \\
5\emph{d}$^{7}$6\emph{p} & $F^{4}(5d,5d)$ & 44375.0 & 473.0 & 2 & 50491.4 & 0.879 & \\
5\emph{d}$^{7}$6\emph{p} & $\alpha(5d)$ & 21.9 & 4.0 & 3 & 0.0 & & \\
... & ... & ... & ... & ... & ... & ... & ... \\
... & $R^{k}$ & & & & & 0.750 & R \\
\enddata
\tablecomments{$^{a}$ Configurations involved in the calculations and their defining Slater parameters with their Hartree-Fock (HFR) and/or least-squares-fitted (LSF) value or their ratio. 
$^{b}$ Uncertainty of each parameter represents their standard deviation. Blank for fixed values. 
$^{c}$ Parameters in each numbered group were linked together with their ratio fixed at the HFR level.
$^{d}$ Comments: F–-The parameters are fixed at given LSF/HFR ratio; R-–All configuration-interaction parameters \emph{R}
$^{k}$ in both even and odd parity sets are fixed at 75\% of their HFR value.}
(This table is available in its entirety in machine-readable form.)
\end{deluxetable*}

\subsection{The HFR+CPOL Calculations} \label{sec:HFR+CPOL}

The physical model used for the HFR+CPOL calculations is based on the consideration of valence-valence correlations within a set of interacting configurations including $5d^8$, $5d^7ns$ ($n$ = 6--8), $5d^7nd$ ($n$ = 6--8), $5d^{6}6s^2$, 5d$^6$6p$^2$, 5d$^6$6d$^2$, 5d$^6$6s6d, 5d$^6$6s7s and 5d$^7$$n$p ($n$ = 6--8), 5d$^7$$n$f ($n$ = 5--6), 5d$^6$6s$n$p ($n$ = 6--7), 5d$^6$6p6d, 5d$^{6}6s5f$ for even and odd parities, respectively. This list of configurations includes a large amount of valence correlation outside of the W-like Au VI ionic core. Therefore, we estimated the core-valence interactions by considering core-polarization contributions with the dipole polarizability, $\alpha_{d}$, equal to 3.62 a$_0^3$ and a cut-off radius, $r_c$, equal to 1.40 a$_0$. As no data are available in the literature, the first parameter was deduced from the extrapolation of $\alpha_d$-values published by \citet{Fraga1976-book} along the tungsten isoelectronic sequence, from W I to Pt V. As for the cut-off radius, it corresponds to the HFR average value $<$$r$$>$ for the outermost core orbital (5d). The semi-empirical LSF approach performed using this HFR+CPOL model gave similar results to those obtained in our HFR calculations with standard deviations of 111 cm$^{-1}$ and 173 cm$^{-1}$ in even and odd parities, respectively. 

\subsection{The MCDHF Calculations} \label{sec:mcdhf}

The MCDHF calculations are based on two multireferences (MR) for even and odd parities including the 5d$^8$, 5d$^7$6s, 5d$^6$6s$^2$ and 5d$^7$6p, 5d$^6$6s6p configurations, respectively. The orbitals were optimized using all the energy levels from each MR. Valence-valence (VV) correlations were estimated by considering single and double (SD) excitations from the MR configurations to 7s, 7p, 7d and 7f orbitals, giving rise to 143441 and 602493 configuration state functions (CSFs) in the even and odd parities, respectively. In this step, only the new orbitals were optimized, the other ones being kept to their values obtained before. Then core-valence (CV) effects were taken into account by considering SD excitations from the 4d, 4f, 5s, 5p core orbitals to the valence orbitals characterizing the MR configurations. This led to 344673 CSFs (even parity) and 623038 CSFs (odd parity). The electric dipole (E1) transitions, allowing us to estimate the MCDHF radiative lifetimes of Au IV levels, were computed by gathering the two parities while the transition probabilities for forbidden lines (M1 and E2) were considered only within the even parity. 

\section{Evaluation of Transition Rates} \label{sec:Tp-eval}

\begin{figure}[t!]
\epsscale{1.16}
\plotone{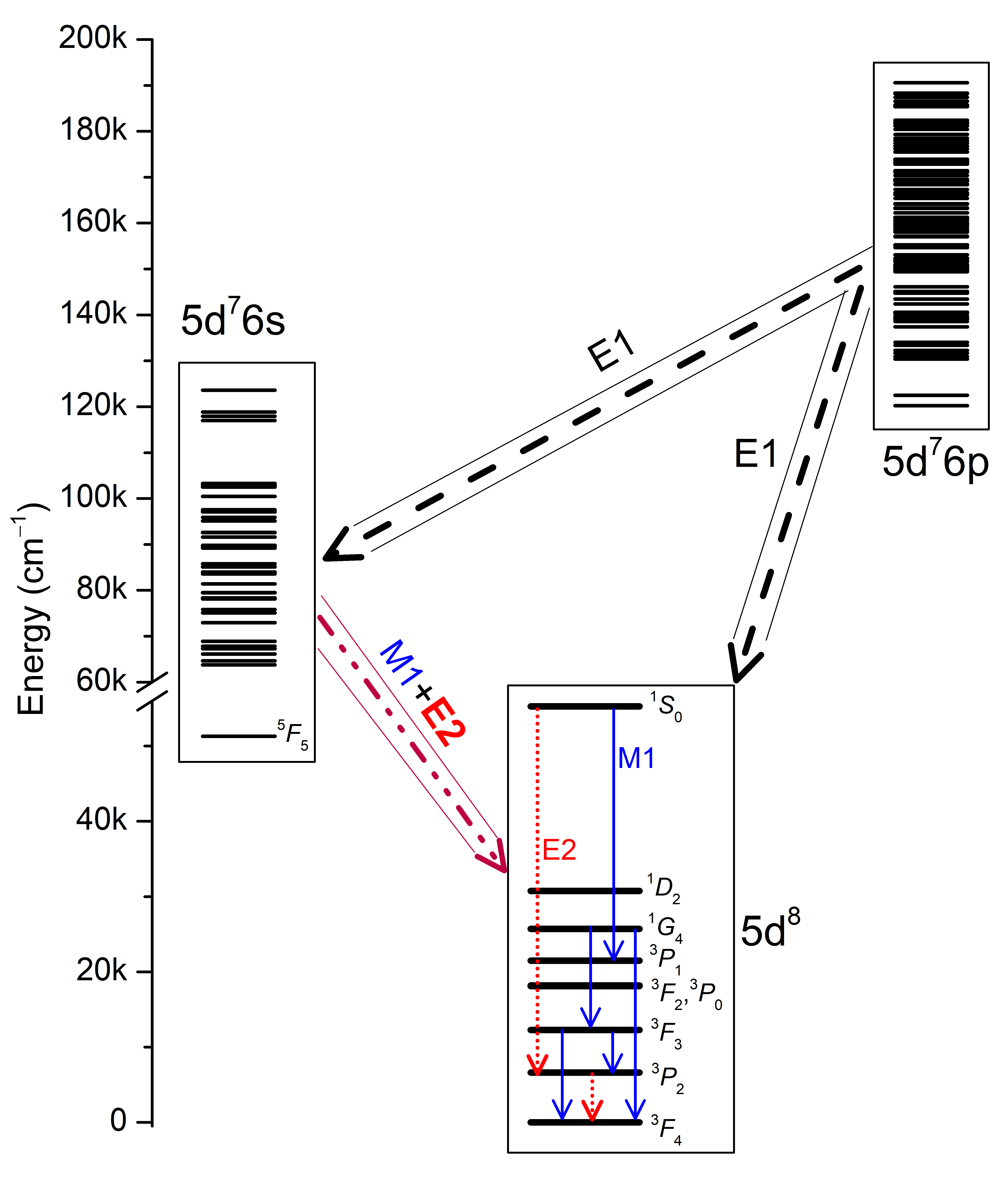}
\caption{Energy level diagram for the known levels of Au IV. Fine structure levels of the ground $5d^{8}$ configuration are shown along with their selected forbidden (M1 and E2) transitions. The transitions within the levels of $5d^{7}6s$ are not shown.}
\label{fig:ELD}
\end{figure}

An energy level diagram for the known levels of Au~IV is illustrated in Figure \ref{fig:ELD} along with its intra- and inter-configuration transitions. In terms of the observations, only the allowed electric dipole E1-types of ($5d^{8}$ + $5d^{7}6s$)--$5d^{7}6p$ transitions are observed (see Table \ref{tab:lines}).
The forbidden M1-type (magnetic dipole) of radiations are dominated in the intra-configuration transitions, for example, within the levels of $5d^{8}$ and $5d^{7}6s$ configurations. However, in case of the $5d^{8}$--$5d^{7}6s$ inter-configuration transitions the transition amplitudes are largely forbidden E2-types (electric quadrupole).

For E1-types, the transitions rates (\emph{A}-values) were obtained from the HFR and HFR+CPOL methods and their comparison are described in section \ref{sec:tp-e1}. On the other hand for forbidden M1- and E2- types the transition parameters including \emph{A}-values were obtained from three different methods (namely: MCDHF, HFR, and HFR+CPOL), and their comparisons are discussed in a separate section \ref{sec:tp-forbide}. 

\subsection{Allowed E1-Type Transitions} \label{sec:tp-e1}

\begin{figure}[b!]
\epsscale{1.15}
\plotone{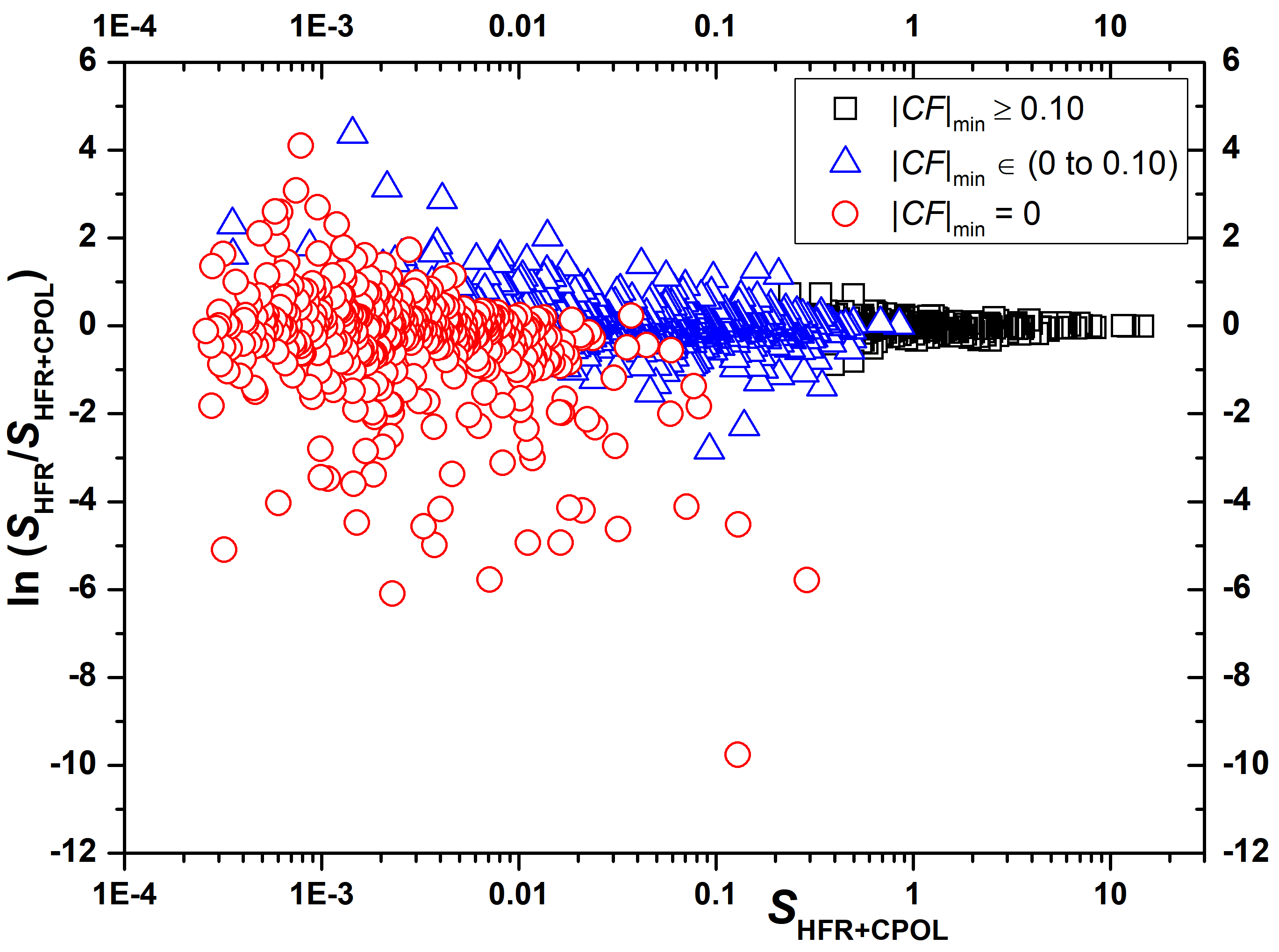}
\caption{Comparison of line strengths (\emph{S}--values) computed within the framework of the HFR and HFR+CPOL calculations for E1-types. For uncertainty evaluations, the entire data was divided into three separate groups depending on their \emph{$|$CF$|$}$_{min}$-values (see text).}
\label{fig:TPE1}
\end{figure}

As discussed briefly above, inclusion of CPOL effects in HFR calculations significantly improves the accuracy of transition rate data (or lifetime, $\tau_{k}$ = 1/$\Sigma$\emph{A}$_{ki}$ data) for Os I and Ir II \citep{Quinet2006-Os1-2,Xu2007Ir1-2}. These authors have shown that their computed lifetime data from the HFR+CPOL calculations agree fairly well (within 30\% and 20\% for Os I and Ir II, respectively) with the experimental lifetimes data (for 10 available levels of the 5\emph{d}$^{7}$6\emph{p} configuration). No experimental lifetime data is available for higher members of Os~I sequence including Au~IV spectrum. Nonetheless, comparison of transition rates from different sources is a prerequisite to estimate the uncertainty of transition rate data \citep{kramida2013critical,haris2014Sn2}. The computed lifetime data {for the levels of $5d^76p$} from the HFR calculations are systematically (about 73 \%) lower to those of the HFR+CPOL ones. Therefore, the HFR lifetime data are scaled to the HFR+CPOL level, and using the scaled lifetime data, the HFR transition rates were re-calculated. For comparison, the transition rates are first converted into their corresponding ``line-strengths'' (\emph{S}-values) using the numerical expression, \emph{S} = \emph{gA}$\lambda^{3}$/2.0261269e+18, where {line strength} \emph{S} (in atomic units, AU), weighted transition rate \emph{gA} (in s$^{-1}$), and wavelength $\lambda$ (in \AA) are expressed in their customary units. The logarithmic difference of two sets of $S$-values (HFR and HFR+CPOL) as a function of \emph{S}$_{HFR+CPOL}$ is plotted, and is shown in Figure \ref{fig:TPE1}. It is evident from Figure \ref{fig:TPE1} that the uncertainty varies with the \emph{S}-values, however, apart from the dependency on magnitude of the S-value it also depends on the absolute value of the cancellation factor \citep{Cowan1981-theory-code}. Both HFR and HFR+CPOL calculations provide the \emph{CF}-factor, thus a minimum value, \emph{$|$CF$|$}$_{min}$ is taken into account. Generally, the \emph{S}-values (or transition rates) with \emph{$|$CF$|$}~$<$~0.10 are considered as unreliable \citep{kramida2013critical}, but we found that the average deviation of \emph{S}-values with 0~$<$~\emph{$|$CF$|$}$_{min}$~$<$~0.10 is about 62 \% and those with \emph{$|$CF$|$}$_{min}$~=~0 deviate more than three orders of magnitude. In fact the lines with \emph{$|$CF$|$}$_{min}$~$\ge$~0.10 show least deviation or they have the higher accuracy. The uncertainty evaluations are made separately for three different cases of \emph{$|$CF$|$}$_{min}$ and the results are summarized in column 2 of Table \ref{tab:tp-code}. It should be noted that these uncertainties represent the statistical (Unc.$_{stat}$) ones for the two data sets, and for a more comprehensive uncertainty estimates, experimental lifetime data and their agreement with the theoretical data are required. Since no experimental lifetime data is available for Au~IV, a conservative estimate of additional (systematic) uncertainty of about 30\% was adopted from the Os I and Ir II lifetime data comparisons as discussed above. To derive the final total uncertainties (Unc.$_{tot}$) for \emph{A}-values, the systematic uncertainty is combined in quadrature with Unc.$_{stat}$, and their corresponding uncertainty codes are provided in Table \ref{tab:lines}. 

\begin{deluxetable}{lcrlcc}[b!]
\tablecaption{Uncertainty Estimate and Code for Transition Rate of Allowed (E1) and Forbidden (M1~+~E2)-Types} \label{tab:tp-code}
\tablewidth{0pt}
\tablehead{
\multicolumn{2}{c}{Unc.$_{stat}$ for E1-type~$^{a}$} & & \multicolumn{3}{c}{Unc.$_{tot}$~$^{b}$} \\
\cline{1-2}\cline{4-6}
\colhead{\emph{S}$_{HFR+CPOL}$, (AU)} & \colhead{Unc.$_{stat}$} & & \colhead{Symbol} & \colhead{in \emph{A}-value} & \colhead{in log(\emph{gf})}
}
\startdata
- & - & & C & $\le$25\% & $\le$0.11 \\
\multicolumn{6}{l}{i) For \emph{$|$CF$|$}$_{min}$~$\ge$~0.10} \\
\cline{1-2}
$\ge$2 & 10\% & & D+ & $\le$40\% & $\le$0.18 \\
$[$0.2, 2) & 17\% & & D+ & $\le$40\% & $\le$0.18 \\
\cline{1-2}
\multicolumn{6}{l}{ii) For \emph{$|$CF$|$}$_{min}$~$<$~0.10} \\
\cline{1-2}
$[$0.2, 0.9) & 25\% & & D+ & $\le$40\% & $\le$0.18 \\
$[$0.02, 0.2) & 35\% & & D & $\le$50\% & $\le$0.24 \\
$[$0.002, 0.02) & 50\% & & E & $>$50\% & $>$0.24 \\
\cline{1-2}
\multicolumn{6}{l}{iii) For \emph{$|$CF$|$}$_{min}$~=~0} \\
\cline{1-2}
$<$0.002 & 300\% & & E & $>$50\% & $>$0.24 \\
\enddata
\tablecomments{$^{a}$ The statistical uncertainty (Unc.$_{stat}$) is estimated from the comparison of two (HFR and HFR+CPOL) datasets. $^{b}$ The total uncertainty (Unc.$_{tot}$) includes an additional 30\% of systematic component, combined in quadrature with the Unc.$_{stat}$ for all E1-type (see Section \ref{sec:tp-e1}). For forbidden (M1~+~E2)-types the Unc.$_{tot}$ determined from the comparison of different dataset (see Section \ref{sec:tp-forbide}).}
\end{deluxetable}

\subsection{Forbidden M1- and E2-Type Transitions} \label{sec:tp-forbide}

\begin{figure*}
\epsscale{1.15}
\plotone{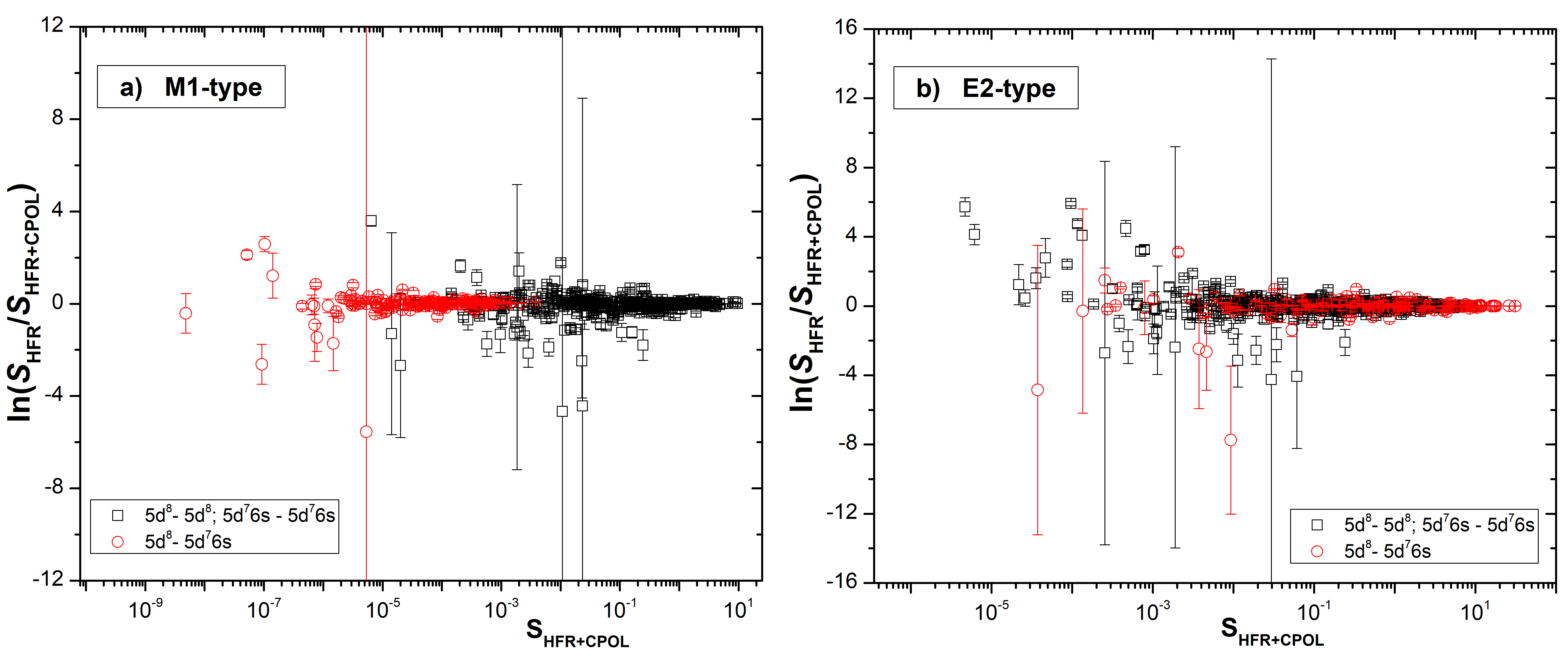}
\caption{Comparison of line strengths (\emph{S}-values) computed within the framework of the HFR and HFR+CPOL methods for the intra- and inter-configuration transitions (see text).}
\label{fig:M1E2-HF}
\end{figure*}

\begin{figure*}
\epsscale{1.15}
\plotone{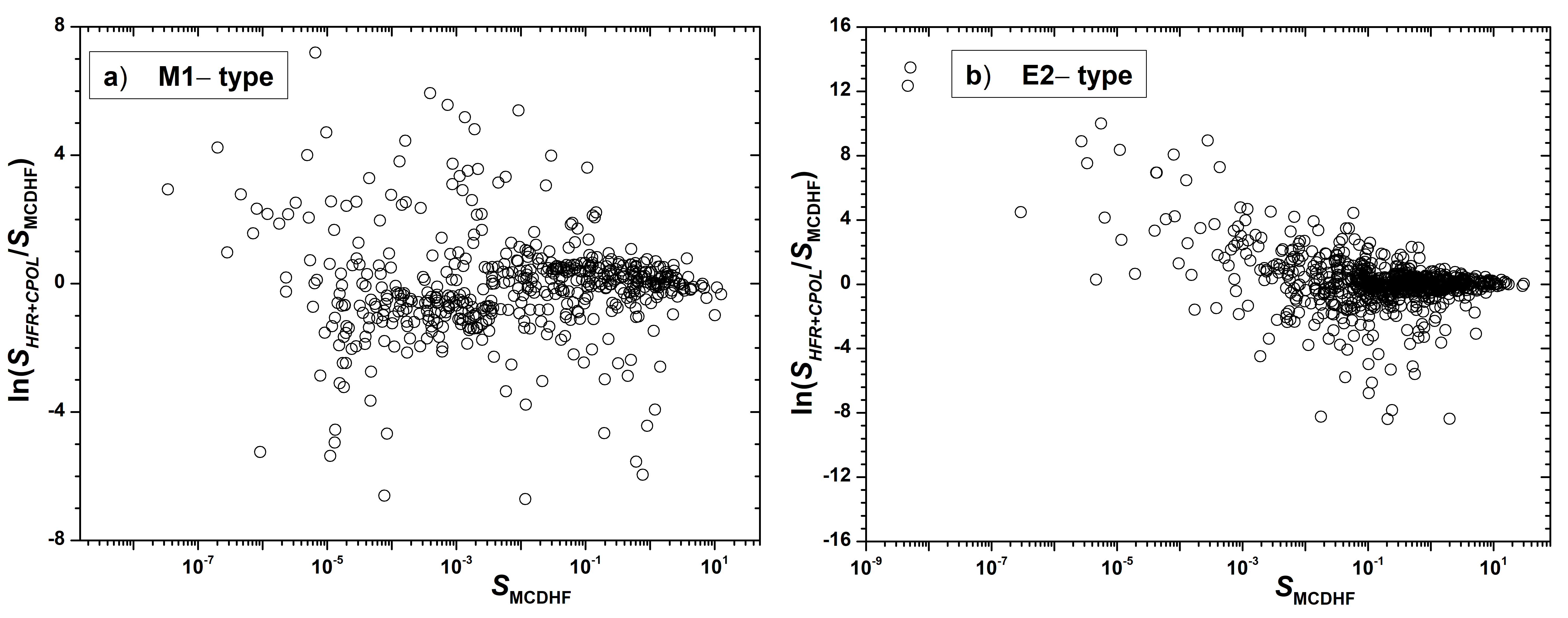}
\caption{Logarithmic difference between the line-strengths (\emph{S}) from MCDHF and HFR+CPOL calculations plotted as a function of \emph{S}$_{MCDHF}$-values (see text).}
\label{fig:M1E2-MCD}
\end{figure*}

In this section, a comparative evaluation of the \emph{A}-values for the forbidden transitions, within the levels of $5d^{8}$ and $5d^{7}6s$ and those of $5d^{8}$--$5d^{7}6s$, from three different computational approaches, \emph{viz.} MCDHF, HFR, and HFR+CPOL has been presented. For the comparison purposes, the initially obtained \emph{A}-values from these sources were first normalized or rescaled with respect to the experimental transition energies, which are computed from Table \ref{tab:energy}. These \emph{A}-values were also used to compute the radiative lifetimes for the levels of $5d^{8}$ and $5d^{7}6s$ configurations, and these lifetimes are given in Table \ref{tab:lifetime}.

\begin{deluxetable}{lcccc}[ht!]
\tablecaption{Radiative Lifetimes $\tau$ (in ms) for the Levels of $5d^{8}$ and $5d^{7}6s$ in Au IV} 
\label{tab:lifetime}
\tablewidth{0pt}
\tablehead{
\colhead{Level $^{a}$} & 
\colhead{$\Delta$E$_{O-M}$ $^{b}$} &
\multicolumn{2}{c}{MCDHF $^{c}$} &
\colhead{HFR+CPOL $^{c}$}
\\
\cline{3-4}
\colhead{} & \colhead{(cm$^{-1}$)} &
\colhead{$\tau^{v}$} & \colhead{$\tau^{l}$} &
\colhead{$\tau$}
}
\startdata
5\emph{d}$^{8}$ $^{3}$\emph{F}$_{4}$ & 0 & .. & .. & .. \\
5\emph{d}$^{8}$ $^{3}$\emph{P}$_{2}$ & -1157 & 2284000 & 1136000 & 891000 \\
5\emph{d}$^{8}$ $^{3}$\emph{F}$_{3}$ & 810 & 21.00 & 21.00 & 21.27 \\
5\emph{d}$^{8}$ $^{3}$\emph{F}$_{2}$ & -307 & 38.1 & 37.9 & 37.9 \\
5\emph{d}$^{8}$ $^{3}$\emph{P}$_{0}$ & -2101 & 930000 & 12130 & 11430 \\
5\emph{d}$^{8}$ $^{3}$\emph{P}$_{1}$ & -1491 & 44.1 & 44.1 & 36.5 \\
5\emph{d}$^{8}$ $^{1}$\emph{G}$_{4}$ & -2330 & 45.8 & 46.1 & 34.4 \\
5\emph{d}$^{8}$ $^{1}$\emph{D}$_{2}$ & 391 & 10.57 & 10.54 & 9.12 \\
5\emph{d}$^{7}$($^{4}$\emph{F})6\emph{s} $^{5}$\emph{F}$_{5}$ & 2426 & 65.2 & 117.8 & 105.8 \\
5\emph{d}$^{8}$ $^{1}$\emph{S}$_{0}$ & -2001 & 3.21 & 2.90 & 2.60 \\
5\emph{d}$^{7}$($^{4}$\emph{F})6\emph{s} $^{5}$\emph{F}$_{4}$ & 2310 & 8.41 & 12.59 & 8.47 \\
5\emph{d}$^{7}$($^{4}$\emph{F})6\emph{s} $^{5}$\emph{F}$_{3}$ & 2598 & 28.9 & 31.2 & 26.8 \\
5\emph{d}$^{7}$($^{4}$\emph{P})6\emph{s} $^{5}$\emph{P}$_{2}$ & 1756 & 9.80 & 10.59 & 7.41 \\
5\emph{d}$^{7}$($^{4}$\emph{F})6\emph{s} $^{5}$\emph{F}$_{1}$ & 1937 & 9.21 & 11.89 & 8.58 \\
5\emph{d}$^{7}$($^{4}$\emph{F})6\emph{s} $^{5}$\emph{F}$_{2}$ & 1289 & 7.45 & 8.99 & 14.70 \\
5\emph{d}$^{7}$($^{4}$\emph{F})6\emph{s} $^{3}$\emph{F}$_{4}$ & 1420 & 1.147 & 1.595 & 1.684 \\
5\emph{d}$^{7}$($^{4}$\emph{P})6\emph{s} $^{5}$\emph{P}$_{3}$ & 1142 & 7.69 & 9.15 & 9.58 \\
5\emph{d}$^{7}$($^{4}$\emph{P})6\emph{s} $^{5}$\emph{P}$_{1}$ & 1219 & 6.06 & 7.86 & 9.31 \\
5\emph{d}$^{7}$($^{2}$\emph{G})6\emph{s} $^{3}$\emph{G}$_{5}$ & 754 & 1.812 & 2.456 & 2.211 \\
5\emph{d}$^{7}$($^{4}$\emph{F})6\emph{s} $^{3}$\emph{F}$_{3}$ & 1301 & 1.116 & 1.443 & 1.304 \\
5\emph{d}$^{7}$($^{4}$\emph{F})6\emph{s} $^{3}$\emph{F}$_{2}$ & 1094 & 1.142 & 1.502 & 1.373 \\
5\emph{d}$^{7}$($^{2}$\emph{G})6\emph{s} $^{3}$\emph{G}$_{4}$ & 630 & 1.840 & 2.59 & 2.130 \\
5\emph{d}$^{7}$($^{2}$\emph{P})6\emph{s} $^{1}$\emph{P}$_{1}$ & 640 & 1.139 & 1.404 & 1.315 \\
5\emph{d}$^{7}$($^{4}$\emph{P})6\emph{s} $^{3}$\emph{P}$_{2}$ & 511 & 0.839 & 0.969 & 0.900 \\
5\emph{d}$^{7}$($^{2}$\emph{H})6\emph{s} $^{3}$\emph{H}$_{6}$ & -536 & 1.394 & 2.097 & 1.779 \\
5\emph{d}$^{7}$($^{2}$\emph{P})6\emph{s} $^{3}$\emph{P}$_{0}$ & 471 & 1.493 & 1.934 & 1.475 \\
5\emph{d}$^{7}$($^{2}$\emph{G})6\emph{s} $^{3}$\emph{G}$_{3}$ & 919 & 2.207 & 2.92 & 2.84 \\
5\emph{d}$^{7}$($^{2}$\emph{H})6\emph{s} $^{3}$\emph{H}$_{5}$ & -634 & 1.869 & 2.92 & 2.477 \\
5\emph{d}$^{7}$($^{2}$\emph{G})6\emph{s} $^{1}$\emph{G}$_{4}$ & 5 & 1.196 & 1.675 & 1.482 \\
5\emph{d}$^{7}$($^{2}$\emph{Db})6\emph{s} $^{3}$\emph{D}$_{3}$ & 1934 & 1.045 & 1.205 & 1.117 \\
5\emph{d}$^{7}$($^{2}$\emph{Db})6\emph{s} $^{3}$\emph{D}$_{2}$ & 858 & 0.833 & 1.049 & 1.080 \\
5\emph{d}$^{7}$($^{4}$\emph{P})6\emph{s} $^{3}$\emph{P}$_{1}$ & 1216 & 0.879 & 0.969 & 0.987 \\
5\emph{d}$^{7}$($^{2}$\emph{F})6\emph{s} $^{3}$\emph{F}$_{2}$ & 845 & 1.036 & 1.164 & 1.560 \\
5\emph{d}$^{7}$($^{2}$\emph{P})6\emph{s} $^{3}$\emph{P}$_{2}$ & -1311 & 1.807 & 2.352 & 1.142 \\
5\emph{d}$^{7}$($^{2}$\emph{F})6\emph{s} $^{3}$\emph{F}$_{3}$ & -2262 & 1.250 & 1.566 & 1.242 \\
5\emph{d}$^{7}$($^{2}$\emph{P})6\emph{s} $^{3}$\emph{P}$_{1}$ & 1190 & 1.216 & 1.432 & 1.399 \\
5\emph{d}$^{7}$($^{2}$\emph{H})6\emph{s} $^{3}$\emph{H}$_{4}$ & 286 & 1.358 & 1.861 & 1.733 \\
5\emph{d}$^{7}$($^{2}$\emph{H})6\emph{s} $^{1}$\emph{H}$_{5}$ & -273 & 0.870 & 1.250 & 1.095 \\
5\emph{d}$^{7}$($^{4}$\emph{P})6\emph{s} $^{3}$\emph{P}$_{0}$ & 995 & 0.537 & 0.608 & 0.603 \\
5\emph{d}$^{7}$($^{2}$\emph{F})6\emph{s} $^{3}$\emph{F}$_{4}$ & -926 & 1.205 & 1.538 & 1.181 \\
5\emph{d}$^{7}$($^{2}$\emph{Db})6\emph{s} $^{1}$\emph{D}$_{2}$ & -531 & 0.780 & 0.915 & 0.788 \\
5\emph{d}$^{7}$($^{2}$\emph{Db})6\emph{s} $^{3}$\emph{D}$_{1}$ & -482 & 0.846 & 0.996 & 0.836 \\
5\emph{d}$^{7}$($^{2}$\emph{F})6\emph{s} $^{1}$\emph{F}$_{3}$ & -1945 & 0.661 & 0.787 & 0.739 \\
5\emph{d}$^{7}$($^{2}$\emph{Da})6\emph{s} $^{3}$\emph{D}$_{1}$ & 250 & 0.678 & 0.661 & 0.566 \\
5\emph{d}$^{7}$($^{2}$\emph{Da})6\emph{s} $^{3}$\emph{D}$_{2}$ & -162 & 0.638 & 0.647 & 0.546 \\
5\emph{d}$^{7}$($^{2}$\emph{Da})6\emph{s} $^{3}$\emph{D}$_{3}$ & -2517 & 0.633 & 0.648 & 0.525 \\
5\emph{d}$^{7}$($^{2}$\emph{Da})6\emph{s} $^{1}$\emph{D}$_{2}$ & -2292 & 0.425 & 0.432 & 0.374 \\
\enddata
\tablecomments{$^{a}$ The levels are from Table \ref{tab:energy}. $^{b}$ Difference between the observed energy level value and its corresponding MCDHF value in $cm^{-1}$. $^{c}$ The MCDHF (in velocity and lengths forms) and HFR+CPOL radiative lifetimes (in millisecond) for the levels were determined using the scaled \emph{A}-values. The scaling were carried out with the help of transition energies computed from Table \ref{tab:energy} (see text).}
\end{deluxetable}

The evaluations were started with the comparison of HFR and HFR+CPOL results for M1- and E2- type of transitions which is shown in Figure \ref{fig:M1E2-HF}. The error bars in this figure represents the internal uncertainties of the \emph{S}-values obtained from the HFR calculations, which has been evaluated using the Monte Carlo technique \citep{Kramida2014-monte-carlo}. A total of 20 trials were made to estimate the uncertainties (of \emph{A}- or \emph{S}-values) for 537 M1- and 777 E2- types of transitions. About 835 transitions have their uncertainty within 3\%, 238 transitions have 7\%, 132 have 18\%, 61 transitions have uncertainties varying between 25\% to 50\%, and 48 transitions have more than 50\% (half of them have 2 or 3 orders of magnitude deviation). In general, an excellent agreement (within 0.33 dex) is observed for HFR and HFR+CPOL data sets:- as expected the strong lines with \emph{S} (in AU) $\geq$ 1 agreed within 10\% of uncertainty, 18\% for the lines with \emph{S} $\in$ [0.2, 1), 25\% for \emph{S} $\in$ [0.02, 0.2), 40\% for \emph{S} $\in$ [0.0001, 0.02), and the remaining weak lines are deviated by more than 50\%.

However, in order to obtain more reliable estimates, the \emph{S}-values from HFR and HFR+CPOL were further compared with the MCDHF ones. For M1-types of transitions, the \emph{S}$_{MCDHF}$-values agree within 0.79 dex and 0.68 dex for \emph{S}$_{HFR}$ and \emph{S}$_{HFR+CPOL}$, respectively. In the case of E2-types, the \emph{S}$_{MCDHF}$-values are available in the length and velocity (Babushkin and Coulomb gauges, respectively) forms. Comparison of these two forms reflects the internal accuracy of the computation, which is generally given in terms of uncertainty indicator $\delta$\emph{T} defined as, $\delta$\emph{T}~=~$|$\emph{A}$^{l}$--\emph{A}$^{v}|$/max(\emph{A}$^{l}$, \emph{A}$^{v}$), where \emph{A}$^{l}$ and \emph{A}$^{v}$ are the \emph{A}$_{ki}$ values calculated in length and velocity gauges, respectively. Note that a similar indicator, $\delta$\emph{S}, can also be computed from their corresponding \emph{S}-values. However, using $\delta$\emph{T} or $\delta$\emph{S} for computation of uncertainty is normally discouraged, since the presence of `max' term in the denominator underestimates the uncertainty \citep{Kramida2022-FeVII}. Instead a reliable uncertainty estimate can be obtained from the \emph{dS} comparisons, which are made as a function of \emph{S}-values (see section \ref{sec:tp-e1}). For the present case, the global disagreement, between length and velocity forms, for 777 E2-type of transitions is within 1.22 dex or three orders of magnitude. However, this large deviation does not necessarily means the entire computational accuracy is poorer, the stronger \emph{S}-values of a certain type of transitions still have better agreement. As E2-types are dominant for inter-configurations transitions, they ought to have better accuracy. This can be confirmed from the gross deviation for the $5d^{8}$--$5d^{7}6s$ transitions, which is $\approx$0.44 dex and $\approx$1.4 dex for the transitions within the levels of $5d^{8}$ and $5d^{7}6s$ configurations. Both length and velocity forms should produce comparably same results with reasonably accurate approximate wavefunctions, yet the velocity forms are not a good choice for small transition energies \citep{Cowan1981-theory-code}. Therefore, in general, length form results are considered to be more accurate than that of velocity form \citep{Hibbert1974oscillator,Fischer2009}, except in the case of Rydberg type of transitions, for which velocity forms could be more appropriate \citep{Papoulia-2019-Velocity-forms}.

For this reason we have {chosen} the length form for the final comparisons. The agreement of \emph{S}$_{MCDHF}$-values for E2-type transitions is $\approx$0.80 dex with \emph{S}$_{HFR}$ as well as with \emph{S}$_{HFR+CPOL}$-values. The comparison plots of M1- and E2-types for HFR+CPOL and MCDHF calculations are shown in Figure \ref{fig:M1E2-MCD}, which represent a fairly good agreement {since their gross deviation was within 0.68 dex and 0.80 dex, respectively}. For M1- type transitions, the strong lines with \emph{S} (in AU) $\geq$ 2 agreed within 25\% of uncertainty, 38\% for the lines with \emph{S} $\in$ [0.6, 2), 50\% for \emph{S} $\in$ [0.1, 0.6), and the remaining weak lines are deviated by more than 50\%. Similarly for E2- transitions, lines with \emph{S} $\geq$ 5 agreed within 23\%, 40\% for the lines with \emph{S} $\in$ [1, 5), 50\% for \emph{S} $\in$ [0.3, 1), and the rest of the weak lines are deviated by more than 50\%. The data on the transition rates (\emph{A}-values) of the M1- and E2- components, obtained from MCDHF, HFR+CPOL, and HFR calculations, along with their accuracy codes and absolute branching fractions are summarized in Table \ref{tab:Forbidden-lines}. The absolute branching fraction, $BF_{ki}^{abs}$ of a spectral line originating from an upper level \emph{k} to a lower level \emph{i} is defined as $BF_{ki}^{abs}$~=~ $\tau_{k}A_{ki}^{tot}$, where $A_{ki}^{tot}$~=~$A_{ki}^{M1}$+$A_{ki}^{E2}$, and is helpful for line identification and confirmation. To derive the effective uncertainty for $A_{ki}^{tot}$ of a spectral line its M1- and E2- component uncertainties can be directly weighted on the basis of their relative branching factor. {It should be noted that our synthesized forbidden spectrum of Au IV falls in the broad wavelength range 809 \AA--6.8 $\mu$m and its numerous lines are within the bandpass of the Near Infrared Spectrograph (NIRSpec; 0.6 to 5 $\mu$m) and/or Mid-Infrared Instrument (MIRI; 5 to 28 $\mu$m) of the JWST. However, only a few selected strong lines of Au IV will be detected by JWST at its maximum resolution R~=~3000, and all those strong lines were already predicted by \citet{Taghadomi-2022-Au4-Os1}. Further, \citeauthor{Taghadomi-2022-Au4-Os1} reported transition parameters for 14 M1-type of lines between the levels of $5d^8$, which we accessed from their opacity database \citep{Stancil-2022-Opac-database}. However, our comparison shows that their disagreement was about 200\% with our MCDHF and HFR+CPOL and/or HFR results. On the other hand, the agreement between our MCDHF and HFR+CPOL or HFR values were within 20\% for these lines. This indicates higher computational accuracy for our results.}\\

Finally, the lifetime data computed within the framework of HFR, HFR+CPOL and MCDHF calculations (see Table \ref{tab:lifetime}) were compared. The very close agreement, within 4\%, between lifetimes from the HFR and HFR+CPOL methods should not be overlooked. Nevertheless, the HFR+CPOL results are preferred because they are more extensive than the other. In MCDHF framework, the lifetime data are provided in both length and velocity forms. For the present case, their disagreement is within 30\% on average, excluding those of $5d^{8}$~$^{3}P_{0}$ level. This indicates that the internal consistency between length and velocity forms, even for strong branches of several transitions, is not very reliable. Most of them are inter-configurations transitions (E2-types) from the $5d^{7}6s$ configuration. Finally, lifetime data from the HFR+CPOL are compared with those of the MCDHF. It was found that the average deviation of $\tau_{HFR+CPOL}$ was 24\% with respect to $\tau^{l}$ (length form) and 32\% with $\tau^{v}$ (velocity form). The radiative lifetime data for the levels of $5d^{8}$ and $5d^{7}6s$ is presented in Table \ref{tab:lifetime}. For the computed lifetime of the uppermost $5d^{7}6s$~$^{1}D_{2}$ level, at 123610 cm$^{-1}$, its $A_{ki}^{E1}$ to two lower levels of $5d^{7}6p$ is not included. These transitions are allowed but are of high-spin changing inter-combination types, and hence they are very weak.\\

\begin{rotatetable*}
\centerwidetable
\begin{deluxetable*}{lllllcllrlclll}
\tablecaption{Transition Rates for Forbidden M1 and E2 Transitions of Astrophysical Interest in Au IV}
\label{tab:Forbidden-lines}
\tablewidth{800pt}
\tabletypesize{\scriptsize}
\movetabledown=1in
\tablehead{
\colhead{$\lambda_{Ritz}$ $^{a}$} & \colhead{Unc.$^{a}$} & 
\colhead{Lower~Level$^{b}$} & \colhead{Upper~Level$^{b}$} & 
\multicolumn{4}{c}{\emph{A} (in s$^{-1}$) for M1-component $^{c}$} &
\colhead{} &
\multicolumn{4}{c}{\emph{A} (in s$^{-1}$) for E2-component $^{c}$} & 
\colhead{BF$_{abs}$ $^{e}$}
\\
\cline{5-8}
\cline{10-13}
\colhead{\AA} & \colhead{\AA} &
\colhead{} & \colhead{} &
\colhead{MCDHF} & \colhead{HFR+CPOL} & \colhead{HFR} & \colhead{Acc. $^{d}$} &
\colhead{} &
\colhead{MCDHF} & \colhead{HFR+CPOL} & \colhead{HFR} & \colhead{Acc. $^{d}$} &
\colhead{} 
} 
\startdata
809.0 & 1.3 & 5\emph{d}$^{8}$ $^{3}$\emph{F}$_{4}$ & 5\emph{d}$^{7}$($^{2}$\emph{Da})6\emph{s} $^{1}$\emph{D}$_{2}$ & & & & & & 1.83e+01 & 1.80e+01 & 2.14e+01 & E & 0.0079 \\
841.5 & 1.4 & 5\emph{d}$^{8}$ $^{3}$\emph{F}$_{4}$ & 5\emph{d}$^{7}$($^{2}$\emph{Da})6\emph{s} $^{3}$\emph{D}$_{3}$ & 2.50e-01 & 3.81e-02 & 5.17e-02 & E & & 1.85e+02 & 1.40e+02 & 1.91e+02 & D & 0.1197 \\
848.1 & 1.4 & 5\emph{d}$^{8}$ $^{3}$\emph{F}$_{4}$ & 5\emph{d}$^{7}$($^{2}$\emph{Da})6\emph{s} $^{3}$\emph{D}$_{2}$ & & & & & & 3.70e+01 & 2.30e+01 & 3.36e+01 & E & 0.0239 \\
854.8 & 1.5 & 5\emph{d}$^{8}$ $^{3}$\emph{P}$_{2}$ & 5\emph{d}$^{7}$($^{2}$\emph{Da})6\emph{s} $^{1}$\emph{D}$_{2}$ & 1.69e-01 & 1.42e-02 & 1.02e-02 & E & & 2.31e+01 & 5.89e+00 & 1.07e+01 & E & 0.0100 \\
... & ... & ... & ... & ... & ... & ... & ... & & ... & ... & ... & ... & ... \\
1659.480 & 0.004 & 5\emph{d}$^{8}$ $^{3}$\emph{F}$_{2}$ & 5\emph{d}$^{7}$($^{2}$\emph{G})6\emph{s} $^{3}$\emph{G}$_{4}$ & & & & & & 2.88e+01 & 4.22e+01 & 4.16e+01 & D+ & 0.0745 \\
1664.022 & 0.005 & 5\emph{d}$^{8}$ $^{1}$\emph{G}$_{4}$ & 5\emph{d}$^{7}$($^{2}$\emph{G})6\emph{s} $^{1}$\emph{G}$_{4}$ & 2.71e-01 & 6.05e-02 & 6.15e-02 & E & & 1.23e+02 & 1.15e+02 & 1.15e+02 & C & 0.2073 \\
1665.680 & 0.004 & 5\emph{d}$^{8}$ $^{3}$\emph{F}$_{2}$ & 5\emph{d}$^{7}$($^{4}$\emph{F})6\emph{s} $^{3}$\emph{F}$_{2}$ & 7.08e-01 & 8.50e-02 & 1.12e-01 & E & & 2.54e+01 & 2.38e+01 & 2.74e+01 & D+ & 0.0393 \\
1668.021 & 0.010 & 5\emph{d}$^{8}$ $^{3}$\emph{P}$_{0}$ & 5\emph{d}$^{7}$($^{4}$\emph{F})6\emph{s} $^{3}$\emph{F}$_{2}$ & & & & & & 2.97e+01 & 5.38e+01 & 4.27e+01 & D+ & 0.0447 \\
... & ... & ... & ... & ... & ... & ... & ... & & ... & ... & ... & ... & ... \\
3222 & 21 & 5\emph{d}$^{7}$($^{2}$\emph{P})6\emph{s} $^{3}$\emph{P}$_{2}$ & 5\emph{d}$^{7}$($^{2}$\emph{Da})6\emph{s} $^{1}$\emph{D}$_{2}$ & 4.76e+01 & 2.48e+01 & 1.79e+01 & D & & 2.79e-04 & 4.04e-01 & 4.79e-01 & E & 0.0205 \\
3228 & 21 & 5\emph{d}$^{7}$($^{2}$\emph{F})6\emph{s} $^{3}$\emph{F}$_{3}$ & 5\emph{d}$^{7}$($^{2}$\emph{Da})6\emph{s} $^{1}$\emph{D}$_{2}$ & 1.82e+01 & 1.78e+01 & 2.05e+01 & D & & 5.25e-03 & 7.67e-02 & 1.36e-01 & E & 0.0079 \\
3228.761 & 0.013 & 5\emph{d}$^{7}$($^{4}$\emph{F})6\emph{s} $^{5}$\emph{F}$_{4}$ & 5\emph{d}$^{7}$($^{2}$\emph{Db})6\emph{s} $^{3}$\emph{D}$_{3}$ & 5.47e+01 & 4.15e+01 & 4.09e+01 & D & & 2.98e-02 & 2.61e-02 & 2.25e-02 & E & 0.0660 \\
3252.61 & 0.03 & 5\emph{d}$^{8}$ $^{3}$\emph{F}$_{4}$ & 5\emph{d}$^{8}$ $^{1}$\emph{D}$_{2}$ & & & & & & 3.06e-01 & 2.16e-01 & 2.18e-01 & D & 0.0032 \\
... & ... & ... & ... & ... & ... & ... & ... & & ... & ... & ... & ... & ... \\
7455.92 & 0.13 & 5\emph{d}$^{8}$ $^{3}$\emph{F}$_{3}$ & 5\emph{d}$^{8}$ $^{1}$\emph{G}$_{4}$ & 1.76e+00 & 2.38e+00 & 2.37e+00 & D & & 5.06e-06 & 1.02e-04 & 1.02e-04 & E & 0.0814 \\
7467.85 & 0.06 & 5\emph{d}$^{7}$($^{4}$\emph{F})6\emph{s} $^{5}$\emph{F}$_{1}$ & 5\emph{d}$^{7}$($^{4}$\emph{P})6\emph{s} $^{3}$\emph{P}$_{2}$ & 6.98e+00 & 1.00e+01 & 9.56e+00 & D & & 3.05e-03 & 2.79e-03 & 3.27e-03 & D & 0.0068 \\
7467.93 & 0.04 & 5\emph{d}$^{7}$($^{4}$\emph{F})6\emph{s} $^{3}$\emph{F}$_{3}$ & 5\emph{d}$^{7}$($^{2}$\emph{Db})6\emph{s} $^{3}$\emph{D}$_{3}$ & 8.11e+00 & 9.29e+00 & 8.61e+00 & D+ & & 3.40e-05 & 2.71e-05 & 1.37e-05 & E & 0.0098 \\
7484.56 & 0.06 & 5\emph{d}$^{7}$($^{2}$\emph{Db})6\emph{s} $^{3}$\emph{D}$_{2}$ & 5\emph{d}$^{7}$($^{2}$\emph{Db})6\emph{s} $^{3}$\emph{D}$_{1}$ & 1.78e+01 & 1.60e+01 & 1.71e+01 & D+ & & 2.81e-03 & 3.07e-03 & 3.20e-03 & E & 0.0177 \\
... & ... & ... & ... & ... & ... & ... & ... & & ... & ... & ... & ... & ... \\
12755.91 & 0.17 & 5\emph{d}$^{7}$($^{2}$\emph{P})6\emph{s} $^{3}$\emph{P}$_{1}$ & 5\emph{d}$^{7}$($^{2}$\emph{Db})6\emph{s} $^{3}$\emph{D}$_{1}$ & 9.01e-01 & 7.97e-01 & 7.71e-01 & D & & 3.25e-04 & 9.64e-04 & 7.92e-04 & E & 0.0009 \\
12767.77 & 0.12 & 5\emph{d}$^{7}$($^{2}$\emph{F})6\emph{s} $^{3}$\emph{F}$_{3}$ & 5\emph{d}$^{7}$($^{2}$\emph{F})6\emph{s} $^{3}$\emph{F}$_{4}$ & 6.08e+00 & 4.99e+00 & 5.19e+00 & C & & 7.37e-04 & 7.65e-04 & 6.28e-04 & D+ & 0.0094 \\
12900 & 300 & 5\emph{d}$^{7}$($^{4}$\emph{P})6\emph{s} $^{3}$\emph{P}$_{1}$ & 5\emph{d}$^{7}$($^{4}$\emph{P})6\emph{s} $^{3}$\emph{P}$_{0}$ & 1.54e+01 & 1.46e+01 & 1.46e+01 & D+ & & & & & & 0.0094 \\
13049.56 & 0.14 & 5\emph{d}$^{7}$($^{4}$\emph{F})6\emph{s} $^{3}$\emph{F}$_{2}$ & 5\emph{d}$^{7}$($^{2}$\emph{G})6\emph{s} $^{1}$\emph{G}$_{4}$ & & & & & & 1.01e-04 & 5.93e-05 & 5.19e-05 & D & 0.0000 \\
... & ... & ... & ... & ... & ... & ... & ... & & ... & ... & ... & ... & ... \\
452590 & 180 & 5\emph{d}$^{7}$($^{2}$\emph{Db})6\emph{s} $^{3}$\emph{D}$_{2}$ & 5\emph{d}$^{7}$($^{4}$\emph{P})6\emph{s} $^{3}$\emph{P}$_{1}$ & 2.22e-05 & 5.68e-05 & 5.98e-05 & E & & 1.43e-11 & 1.57e-11 & 1.32e-11 & D & 0.0000 \\
1187000 & 6000 & 5\emph{d}$^{8}$ $^{3}$\emph{F}$_{2}$ & 5\emph{d}$^{8}$ $^{3}$\emph{P}$_{0}$ & & & & & & 6.66e-15 & & 2.89e-14 & & 0.0000 \\
1607700 & 2000 & 5\emph{d}$^{7}$($^{2}$\emph{P})6\emph{s} $^{1}$\emph{P}$_{1}$ & 5\emph{d}$^{7}$($^{4}$\emph{P})6\emph{s} $^{3}$\emph{P}$_{2}$ & 9.01e-08 & & & & & 6.30e-14 & & & & 0.0000 \\
1797300 & 2300 & 5\emph{d}$^{7}$($^{2}$\emph{P})6\emph{s} $^{3}$\emph{P}$_{2}$ & 5\emph{d}$^{7}$($^{2}$\emph{F})6\emph{s} $^{3}$\emph{F}$_{3}$ & 3.20e-06 & & & & & 1.32e-15 & & & & 0.0000 \\
\enddata
\tablecomments{\\$^{a}$ Ritz wavelength and its uncertainty. The wavelengths are in standard air for $\lambda$ = (2000--20000) \AA~ region and in vacuum outside this region. Conversion between air and vacuum was made with the five-parameter formula of \citet{Peck1972index}. Wavelength uncertainties are determined in the level optimization procedure (see Section \ref{sec:optimization}).\\
$^{b}$ Level designations and their energies from Table \ref{tab:energy}. Two columns for energies are dropped from this condensed table.\\
$^{c}$ The scaled \emph{A}-values (in s$^{-1}$) for M1- and E2- components from MCDHF, HFR+CPOL, and HFR calculations. The scaling were carried out with the help of transition energies computed from Table \ref{tab:energy}.\\
$^{d}$ Accuracy code of the \emph{A}-value is explained in Table \ref{tab:tp-code}.\\
$^{e}$ Absolute branching fraction for the line is calculated from the MCDHF-values.}
(This table is available in its entirety in machine-readable form.)
\end{deluxetable*}
\end{rotatetable*}

{\section{Transition Data of A\lowercase{u} IV for the Calculation Opacity in Kilonova Spectra}\label{sec:Opacity}}

{As briefly mentioned in the introduction, \citet{Tanaka-2020-KNe-opac} and \citet{Taghadomi-2022-Au4-Os1} provided large scale atomic data for Au IV with specific application to compute the opacity in the kilonova ejecta. On the other hand, our results in Tables \ref{tab:lines}, \ref{tab:energy}, \& \ref{tab:Forbidden-lines} describe accurate Ritz wavelengths along with their transition probabilities, optimized energy levels with their uncertainties for $5d^8$, $5d^76s$ and $5d^76p$ configurations, and accurately computed line parameters for several forbidden transitions for levels of $5d^8$ and $5d^76s$ configurations. The accurate atomic data are important for abundance estimates of r-process elements in kilonova ejecta~\citep{Vieira-2023-Abund-KNe}. In their HULLAC code calculations \citep{Tanaka-2020-KNe-opac, Kato-2021-Opac-database} included a limited set of configurations; namely, $5d^8$, $5d^7nl$($n\ge7$, $l = s, p$), $5d^76d$, and $5d^66s^2$, however, they generated about thirty thousand E1-type transitions for their possible 507 levels. Whereas \citet{Taghadomi-2022-Au4-Os1} have performed the calculations using two different versions of GRASP codes with varying set of configurations for Os I isoelectronic sequence including Au IV. For Au IV spectrum, about seventy thousand E1-type transitions were computed with A-values for 551 levels of $5d^8$, $5d^76s$, $5d^76p$, $5d^66s^2$, and $5d^66s6p$ configurations. The GRASP2K calculations were found converging with smaller set of configurations, and the comparison with experimental energy values was found to be in reasonable agreement for levels of the ground $5d^8$ configuration only. Although among atomic structure codes the GRASP calculations are considered to be more accurate, however, the accuracy of those calculations is still a subject of debate for heavy-elements such as gold with open $d$-shells, and large number of configurations sets are often required to build the internal consistency of computed quantities, for example, comparisons of transition probabilities in length and velocity forms.}\\

{To check the quality of these calculations, we have made a gross comparison for the energy levels computed with the LSF of Cowan's code calculations described in Section \ref{sec:HFR} with those computed by \citet{Taghadomi-2022-Au4-Os1}. Their data was accessed from the opacity database \citep{Stancil-2022-Opac-database}. It has been found that the fractional energy differences vary between 5\% and 15\%, and the gross disagreement for A-values were found to be accurate within two-to-three orders of magnitude for lines in the ($5d^8$+$5d^76s$)--$5d^76p$ transition array. Though the strong transitions with $S\ge10$ agree within 125\% have better agreement, and there were 500\% for $S\in[4, 10)$ but deviations were more than three orders of magnitude for the remaining weak lines. These large deviations even for strong lines could be due to poor computation of the LS-composition vectors in the GRASP calculations. A similar large deviations (200\% for $S\ge10$, 500\% for $S\in[3, 10)$, and more than three orders of magnitudes for other weak lines) were also noticed for the comparisons of HULLAC code results for Au IV by \citet{Tanaka-2020-KNe-opac}, therefore, we argue that our computed radiative parameters for Au IV are expected to be more accurate than those by previous works \citep{Tanaka-2020-KNe-opac, Taghadomi-2022-Au4-Os1} as we considered an extensive sets of almost all interacting configurations in our calculations (see Table \ref{tab:params}). This motivated us to provide large scale data for computing the opacity of Au IV in the kilonova ejecta. Nonetheless, from the HFR calculations described in Section \ref{sec:HFR}, only selected set of configurations, $5d^8$, $5d^7$($6s+7s+6d$), $5d^66s^2$, $5d^7$($6p+7p+5f$), and $5d^66s6p$, for which most of the interacting configurations were accounted for has been considered for the computation of radiative transition parameters. This resulted in computing the radiative line parameters for about ninety five thousand E1-type transitions between 1070 levels of the selected configuration set mentioned above. All those data were supplemented by us \citep{zainab_aashna_2023_7788722}. It should be noted that the LS-composition vector provided for the levels have little physical meaning due to very high configuration mixing but they are kept in the supplementary table (of energy levels) for bookkeeping purpose only.}

\section{Conclusion}\label{sec:conclusion}

A critically evaluated set of data for Au IV ion has been presented in the wavelength region 500--2106~\AA. Theoretical interpretation has been done using Cowan’s suite of codes which works on the formalism of pseudo-relativistic Hartree–Fock method. 981 spectral lines have been used to optimize 139 energy levels of $5d^{8}, 5d^{7}6s$, and $5d^{7}6p$ configurations. All previously reported levels have been confirmed except one $J$ = 2 level of $5d^{7}6p$, which is revised now. Also, three new levels, including the only missing $^{1}S_{0}$ level of the ground configuration $5d^{8}$, have been established. Uncertainties have been provided for all the observed and Ritz wavelengths of the spectral transitions. Intensities of the observed lines have been reduced to a common scale. The linelist of Au IV has also been supplemented with Ritz wavelengths of about 830 possibly observable lines and their \emph{gA}-values. Additionally, Ritz wavelengths and radiative parameters of about 800 forbidden (M1- and E2-type) lines have also been provided. For evaluation of the uncertainty of transition probability data, comparisons have been made between the calculations of HFR, HFR+CPOL and MCDHF methods. The overall agreement was found to be fairly well within the three sets of computations. This enabled us to calculate the radiative lifetimes of $5d^{8}$ and $5d^{7}6s$ levels. {Besides these, large scale atomic data with specific application to compute the opacity of Au IV in the kilonova ejecta have been supplemented in this work.}

\begin{acknowledgments}

A. Tauheed would like to duly acknowledge Dr. Y. N. Joshi (late) for giving the spectral plates of gold. The authors highly thankful to Dr. Alexander Kramida of NIST, Gaithersburg, for providing the updated version of the~\citet{Cowan1981-theory-code} codes and for helpful discussions regrading various computational problems. P. Quinet is Research Director of the Belgian National Fund for Scientific Research (F.R.S.-FNRS) from which the financial support is very appreciated. Part of the atomic calculations in Mons are supported by the Consortium des Equipements de Calcul Intensif (CECI) funded by the F.R.S.-FNRS.

\end{acknowledgments}


\bibliography{aaAu4}{}
\bibliographystyle{aasjournal}

\listofchanges

\end{document}